\documentclass[10pt,aps,twocolumn]{revtex4-2}
\usepackage[latin9]{inputenc}
\usepackage[active]{srcltx}
\usepackage{xcolor}
\usepackage{amsmath}
\usepackage{graphicx}
\usepackage{rotfloat}
\PassOptionsToPackage{normalem}{ulem}
\usepackage{ulem}
\usepackage{calc}
\usepackage{multirow}
\usepackage{cancel} 
\usepackage{comment}
\usepackage{appendix}

\makeatletter

\providecolor{lyxadded}{rgb}{0,0,1}
\providecolor{lyxdeleted}{rgb}{1,0,0}
\DeclareRobustCommand{\lyxsout}[1]{\ifx\\#1\else\sout{#1}\fi}

\usepackage{lineno}
\usepackage{hyperref}

\makeatother

\def\HDplus{HD$^+$}
\def\HTplus{HT$^+$}
\def\DTplus{DT$^+$}
\def\T2plus{T$_2^+$}
\def\D2plus{D$_2^+$}
\def\H2plus{H$_2^+$}

\begin{document}
\title{\textcolor{black}{Determination of a set of fundamental constants from molecular hydrogen ion spectroscopy: a modeling study}}

\author{J.-Ph.\,Karr$^{1,2}$, S.\,Schiller$^3$, V.I.\,Korobov$^{4}$, S.\,Alighanbari{$^3$}}

\affiliation{{$^1$}{Laboratoire Kastler Brossel, Sorbonne Universit\'e, CNRS, ENS-Universit\'e PSL, Coll\`{e}ge de France, 4 place Jussieu, F-75005 Paris, France} \\
{$^2$}{Universit\'e Evry Paris-Saclay, Boulevard Fran\c{c}ois Mitterrand, F-91000 Evry, France} \\
{$^3$}{Mathematisch-Naturwissenschaftliche Fakult\"at, Institut f\"ur Experimentalphysik, Heinrich-Heine-Universit\"at D\"usseldorf, 40225 D\"usseldorf, Germany} \\
{$^4$}{Bogoliubov Laboratory of Theoretical Physics, Joint Institute for Nuclear Research, Dubna 141980, Russia}}

\begin{abstract}
The rovibrational transition frequencies of molecular hydrogen ions (MHI) can be accurately computed using ab initio nonrelativistic quantum electrodynamics. A subset of the fundamental constants are required input. We analyze how, once upcoming ultra-high-accuracy spectroscopy data has been obtained, that subset of constants can be determined with greater accuracy. Our analysis 
shows that under realistic assumptions the uncertainties of the mass ratios of proton, deuteron and triton relative to the electron, and of the triton charge radius can be reduced more than one-hundred-fold compared to today (CODATA 2022). Furthermore, the Rydberg constant, as well as the proton and deuteron charge radii can be determined with uncertainties similar to those of today, but solely using data from electronic systems. The implications are discussed. 
\end{abstract}
\maketitle

\vfill\newpage

\noindent\textbf{Introduction}

\noindent The energy scales of atoms and molecules are mainly determined by the values of the fundamental constants (FCs) $R_\infty$ (Rydberg constant), $\alpha$ (fine-structure constant), nucleus-electron mass ratios $m_n/m_e$, electron and nuclear magnetic moments, and nuclear charge radii $r_n$. There is a continuous community effort towards determining the FCs with increasing accuracy~\cite{Mohr2025}. Their values are adjusted so that theory predictions and experimental data match. Over the decades, sophisticated experimental techniques were developed and applied to obtain increasingly accurate data: laser spectroscopy of electronic hydrogen, muonic hydrogen, and helium, electron spin resonance and mass spectrometry of single electrons and single atomic ions in Penning traps, and atom interferometry. The interpretation of most experiments requires a correspondingly sophisticated evaluation of quantum electrodynamics (QED) effects. 

Periodically, the Committee on Data of the International Science Council (CODATA) reviews the available experimental measurements and the theory of the studied systems and recommends updated FC values and uncertainties, based on a global least-squares adjustment (LSA). The 2022 CODATA recommendation~\cite{Mohr2025} included for the first time spectroscopy data and theory of one molecular hydrogen ion (MHI), \HDplus. This occurred as a result of a decades-long development of the ab initio theory of this three-body system, whose transition frequencies $f_i^{\rm theor}$ can today be computed with estimated fractional theoretical uncertainty of $u(\delta f_i^{\rm theor})/f_i^{\rm theor}\simeq8\times10^{-12}$~\cite{Korobov2021} ($u$ is the absolute uncertainty). Experimental measurements of transition frequencies have become feasible with a similar uncertainty level~\cite{Alighanbari2020,Patra2020,Kortunov2021} (see also Tab.\,\ref{tab:Overview of transitions}), thanks to the ability to trap and sympathetically cool the molecules~\cite{Roth2006,Bressel2012}, enabling Doppler-free spectroscopy. Additional high-accuracy results on \HDplus~\cite{Alighanbari2023} and \H2plus~\cite{Alighanbari2024} have more recently been obtained. For a basic introduction to MHI, see e.g.~\cite{Schiller2022}.

With the \HDplus and \H2plus data generated to date, the contribution to FC determination is mostly for the ratio of proton-electron mass and deuteron-electron mass, $m_p/m_e$, $m_d/m_e$. Notably, the value of $m_p/m_e$ from \H2plus~\cite{Alighanbari2024} has lower uncertainty than the earlier determination achieved via combining bound-electron $g$-factor theory and Penning-trap spin and cyclotron resonance experiments~\cite{Koehler2015,Heisse2019}. 

The purpose of this paper is to provide an in-depth analysis of the future potential of precision spectroscopy and theory of \emph{all} MHI (i.e., including also \D2plus, \HTplus, \DTplus, and \T2plus) towards improving the values of the masses of proton, deuteron and triton relative to the electron mass, $R_\infty$, and the charge radii of the named nuclei~\footnote{A first study~\cite{Schiller2024} by us unfortunately contained a numerical error, corrected in~\cite{Schiller2025}}. Obviously, the assumptions entering the analysis will determine the outcome. These assumptions concern the reasonably possible progress in both the experimental and theoretical domain.

We may expect substantial progress in the experimental accuracy, by making use of quantum logic techniques~\cite{Holzapfel2024arXiv} already established for atomic ions. Ultimately, fractional uncertainties in the $10^{-17}$ range appear feasible~\cite{Schiller2014,Karr2014}, far beyond any imaginable future QED uncertainty. The quoted state-of-the-art theory uncertainty $u(\delta f_i^{\rm theor})$ results from unknown high-order-QED contributions~\cite{Korobov2021}. Although work is in progress to reduce the theory uncertainty further~\cite{Kullie2022,Nogueira2023}, one must consider that the QED theory of the much simpler hydrogen atom - which has been researched much more intensely - is only one order more accurate~\footnote{Given the higher theory accuracy for H/D/T (and the availability of the muonic variants $\mu$H, $\mu$D), one may ask whether the present idea could in principle be applied to these systems as well. The answer is affirmative, but in practice it would be very hard to obtain usefully high experimental accuracies for transitions \emph{other} than 1S-2S, due to their generally large natural widths. In this respect HMIs are special.}. In other words, for MHI there are substantial challenges ahead on the theoretical side. 

While it would seem that MHI spectroscopy can improve the accuracy of FCs only as far as the ab initio theory uncertainty is lowered, fortunately this is not so, because the theoretical predictions of the transition frequencies between different pairs of levels in the same isotopologue and even between isotopologues have uncertainties that are strongly correlated. Therefore it might be possible, by selecting suitable transitions and measuring them accurately, to remove to a large extent the (correlated) theoretical uncertainties. FCs and other parameters (e.g. upper limits for beyond-standard-model (BSM) forces) could then be determined with uncertainties much lower than what is possible if only a single or a few transitions are measured. The six isotopologues with their literally tens of thousands of rovibrational transitions of high line quality provide a huge ``reservoir" of transitions. We recently analyzed this possibility~\cite{Schiller2024,Schiller2025}, extending an earlier analysis~\cite{Karr2016a}. Whereas many illustrative examples were considered in~\cite{Schiller2024,Schiller2025}, a thorough optimization of the selected transitions was not performed. 

In the present work, we systematically searched for the best choices within a moderately large set
of experimentally accessible transitions, and show that the FC determinations can be greatly improved by such a careful selection. We furthermore extend our analysis to a scenario where an LSA of FCs is performed on transition frequencies measured on five or all six isotopologues. 

The tritium-bearing isotopologues (T-MHI), \HTplus, \DTplus, \T2plus, have never been studied with laser or THz spectroscopy, while \D2plus\, has not yet been studied with sufficiently high accuracy 
\cite{Carrington1988a,Carrington1989a,Carrington1989b,Carrington1993a,Cruse2008,Beyer2017,Ando2018}.However, high-resolution spectroscopy of these molecules is in principle feasible. Involving the T-MHI gives access to the charge radius $r_t$ of the triton ($t$) and to its mass $m_t$. Measuring accurately $r_t$ is motivated by a precision test of a prediction of the chiral effective field theory of nuclei (for details, see Supplemental Material (SM), sec.~\ref{Appendix - triton charge radius}). The triton mass - a quantity of importance in neutrino mass determination - is known accurately from Penning trap mass spectroscopy and an independent determination using laser spectroscopy would allow a stringent cross-verification of the employed techniques and molecular theory (SM, sec.~\ref{Appendix - triton mass}).

\begin{table*}[htb!]
\begin{centering}
\[
\begin{array}{cccc|ccccccc|ccc}
 i & \text{Mol.} & \text{Transition} & f_i & \multicolumn{7}{c}{\text{Normalized sensitivity to the constant $z$ (kHz)}}  & \multicolumn{3}{c}{\text{Uncertainties (kHz)}} \\
 \text{} & \text{} & 
 %(v,N)\rightarrow (v',N') 
 & \text{(THz)} & m_p/m_e & m_d/m_e & m_t/m_e & c\,R_{\infty } & r_p & r_d & r_t & u(f^{\text{exp}}) & u(\delta f^{\text{theor}}) & \text{Ref.} \\
\hline
 1 & \text{HD}^+ & (0,0\text{)$\to $(}0,1) & 1.3 & -0.015 & -0.0074 & 0 & 0.0014 & -0.00098 & -0.0010 & 0 & 0.015 & 0.019 & \cite{Alighanbari2020} \\
 2 & \text{HD}^+ & (0,0\text{)$\to $(}1,1) & 59 & -0.33 & -0.16 & 0 & 0.064 & -0.026 & -0.028 & 0 & 0.15 & 0.49 & \cite{Kortunov2021} \\
 3 & \text{HD}^+ & (0,0\text{)$\to $(}5,1) & 260 & -1.3 & -0.63 & 0 & 0.28 & -0.11 & -0.12 & 0 & 0.6 & 2.1 & \cite{Alighanbari2023} \\
 4 & \text{HD}^+ & (0,3\text{)$\to $(}9,3) & 415 & -1.7 & -0.84 & 0 & 0.45 & -0.17 & -0.18 & 0 & 0.45 & 3.2 & \cite{Patra2020} \\
 5 & \text{HD}^+ & (9,1\text{)$\to $(}18,0) & 208 & 0.77 & 0.38 & 0 & 0.23 & -0.087 & -0.069 & 0 & \text{} & 1.5 & \text{} \\
 6 & \text{HD}^+ & (7,1\text{)$\to $(}15,0) & 237 & -0.0029 & -0.0015 & 0 & 0.26 & -0.092 & -0.093 & 0 & \text{} & 1.8 & \text{} \\
 8 & \text{HD}^+ & (5,1\text{)$\to $(}13,0) & 278 & -0.49 & -0.24 & 0 & 0.30 & -0.11 & -0.11 & 0 & \text{} & 2.1 & \text{} \\
 9 & \text{H}_2^+ & (1,0\text{)$\to $(}3,2) & 124 & -0.95 & 0 & 0 & 0.14 & -0.11 & 0 & 0 & 0.95 & 1.0 & \cite{Alighanbari2024} \\
 13 & \text{D}_2^+ & (1,0\text{)$\to $(}3,2) & 91 & 0 & -0.72 & 0 & 0.10 & 0 & -0.085 & 0 & \text{} & 0.75 & \text{} \\
 14 & \text{D}_2^+ & (13,0\text{)$\to $(}15,2) & 49 & 0 & 0.0032 & 0 & 0.053 & 0 & -0.040 & 0 & \text{} & 0.39 & \text{} \\
 18 & \text{HT}^+ & (0,0\text{)$\to $(}4,1) & 203 & -1.2 & 0 & -0.86 & 0.22 & -0.088 & 0 & -24. & \text{} & 1.6 & \text{} \\
 23 & \text{T}_2^+ & (0,0\text{)$\to $(}2,2) & 78 & 0 & 0 & -1.4 & 0.086 & 0 & 0 & -20. & \text{} & 0.66 & \text{} \\
 24 & \text{T}_2^+ & (1,0\text{)$\to $(}3,2) & 76 & 0 & 0 & -1.3 & 0.083 & 0 & 0 & -19. & \text{} & 0.63 & \text{} \\
\hline
 25 & \text{H-D} & \text{isotope shift} & \text{} & -0.023 & 0.012 & 0 & 0.00073 & 1.5 & -1.6 & 0 & 0.015 & 0.34 & \cite{Parthey2010} \\
 26 & \text{H} & \text{1S-2S} & \text{} & 0.023 & 0 & 0 & 2.7 & -1.5 & 0 & 0 & 0.010 & 1.4 & \cite{Parthey2011} \\
\end{array}
\]
\par
\end{centering}
\caption{\label{tab:Overview of transitions}
{\footnotesize 
Subset of transitions considered in this work and their properties. In column 3, the transitions are indicated as $(v,N)\rightarrow (v',N') $, where $v$, $N$ are the vibrational and rotational quantum numbers, respectively. The bottom two lines are transitions of electronic hydrogen. Columns 5 -- 11 contain the normalized sensitivities $A_{ij} u_{2022}(z_j)$ of  various transitions of five MHI to the most relevant fundamental constants $z_j$. Note the change in sign of the mass sensitivities for transition 5 and the near-vanishing for transitions 6 and 14. The theoretical uncertainty refers to the uncertainty of the QED corrections, it does not include Bethe logarithm numerical uncertainties (nor FC uncertainties). The transitions $n=1,\ldots12$ are the same as in our previous paper~\cite{Schiller2024}. The remainder of the transition set considered in this work is found in SM, Tab.~\ref{tab:Overview of remaining transitions}. Only one of the two most recent 1S-2S measurements~\cite{Matveev2013,Parthey2011} is shown.}}
\end{table*}

\vskip .1in
\noindent\textbf{Rovibrational transitions of the MHI}

\noindent Tab.\,\ref{tab:Overview of transitions} and SM~Tab.\,\ref{tab:Overview of remaining transitions} summarize the properties and status of a number of transitions considered in the following. They were selected to have convenient transitions  for current laser technology, i.e.~not-too-small frequencies and transition probabilities. We also include (electronic) hydrogen and deuterium 1S - 2S transitions, which play an important role in our analysis. We show in the table the sensitivities of the transition frequencies to the main FCs relevant to both MHI and H/D, denoted by $z_j$. They are given in normalized form, $A_{ij}\times u_{2022}(z_j)$ for easier interpretation. Here, the sensitivity coefficients $A_{ij}$ are defined as
\begin{equation}
     A_{ij} = \partial f_i(z_1,z_2,\ldots)/ \partial z_j\ ,
\end{equation}
while $u_{2022}(z_j)$ is the current CODATA 2022 uncertainty of $z_j$. The normalized sensitivities can be interpreted as the (partial) uncertainties $u_{z_j}(f_i^{\rm theor})$ of the theoretically computed transition frequencies stemming from today's uncertainty of a particular fundamental constant $z_j$~\footnote{The correlations existing between the FC values may be disregarded for the following consideration.}. As can be seen, the uncertainties $|u_{z_j}(f_i^{\rm theor})|$ of the mass ratios ($j=1,2,3$) are typically comparable to the state-of-art uncertainties of the theoretical transition frequencies, $u(\delta f^{\rm theor}_i)$. In contrast, for $r_t$ the
uncertainties are more than one order larger, while for the remaining FCs $R_\infty$, $r_p$, $r_d$ they are approximately one order smaller. One would then expect that only for $r_t$
(and to a lesser extent, for $m_t/m_e$, see transitions 23-24) an improved determination is possible, provided suitable experiments on T-MHI are performed. However, we shall show that all FCs can be substantially improved, except $r_p$, $r_d$, which have  profited from the results of muonic atom spectroscopy. This prospective improvement requires a 2 -- 3~orders reduction in experimental uncertainties compared to today, but no reduction of $u(\delta f^{\rm theor}_i)$. Today's experimental uncertainties are also shown in Tab.~\ref{tab:Overview of transitions}: they are moderately smaller or equal to the theoretical uncertainties. 

\vskip .1in
\noindent\textbf{Least-squares adjustment} 

\noindent In the following, we simulate the determination of FCs from (hypothetical) MHI spectroscopy data using the linearized LSA procedure described in Appendix E of~\cite{Mohr2000} (see also~\cite{Karr2023}). Briefly, the determination of the set of 5 or 7 FCs $\{z_j\}$ from a set of $n$ measured and calculated transition frequencies is associated with $n$ pairs of ``observational equations''~\cite{Mohr2000}:
\begin{subequations}
\begin{gather}
f_i^{\rm exp} \doteq f_i^{\rm theor} + \delta f_i + \sum_{j} A_{ij} (z_j -z_{j0}) \,, \label{eq:obs-eq-1} \\
\delta f_i^{\rm theor} \doteq \delta f_i \,.  
\end{gather}
\end{subequations}
The dotted equality sign means that the left and right hand sides are not equal in general (since the set of equations is overdetermined) but should agree within estimated uncertainties. $\delta f_i$ is an  additive correction that is introduced to take theoretical uncertainties  into account, and is treated as an adjusted constant. The theoretical value $f_i^{\rm theor}$ is calculated for reference (CODATA 2022) values $z_{j0}$ of the FCs $z_j$. The input datum $\delta f_i^{\rm theor}$ is equal to zero with an uncertainty equal to the theoretical uncertainty $u(\delta f_i^{\text{theor}})$. The last term of Eq.~(\ref{eq:obs-eq-1}) corresponds to the linearized dependence of the $f_i^{\rm theor}$ on the FCs around their reference values. Hydrogen and deuterium atom data (see last lines of Tab.~\ref{tab:Overview of transitions}) are treated in the same way. 

Following \cite{Mohr2000}, the covariance matrix of the LSA solution is
\begin{equation}
G = \left( A^{\rm T} V^{-1} A \right)^{-1} \, \label{eq:LSA}
\end{equation}
where $A$ is the sensitivity matrix constructed from the sensitivity coefficients $A_{ij}$ of the selected transitions, and $V$ the covariance matrix of the input data $\{ f_i^{\rm exp},\delta f_i^{\rm theor} \}$. This matrix depends on the uncertainties  $ u(f_i^{\rm exp})$, $u(\delta f_i^{\rm theor})$, and their correlations (see End Matter for details). We emphasize that in our modeling we do not include any CODATA 2022 information on the FCs of interest. For FCs of less relevance to the MHI, mainly $\alpha$, we take the CODATA 2022 value and may safely ignore its uncertainty~\footnote{In the CODATA procedure, the LSA has to be iterated until convergence, due to the fact that the sensitivity coefficients themselves depend on FC values. Here, we make the assumption that the solutions are not too far from initial reference values, such that the first iteration already gives a very good approximation. Under this assumption, the final uncertainties do not depend on differences between $f_i^{\rm exp}$ and $f_i^{\rm theor}$.}.

The computation of the input parameters of our modeling, i.e. the sensitivity coefficients, 
theoretical uncertainties (both shown in Tab.~\ref{tab:Overview of transitions}) and the latter's correlation coefficients, is explained in the End Matter. Our baseline scenario for the estimation of theoretical uncertainties is Ref.~\cite{Korobov2021}, meaning that no QED theory improvements are assumed. A more advanced scenario is described briefly in the Discussion below. Moreover, we improve our previous treatment~\cite{Schiller2024} by taking estimated numerical uncertainties into account, see End Matter for details.

\vskip .1in
\noindent\textbf{Experimental uncertainties} 

\noindent 
We shall assume the same experimental uncertainty for all transitions, both pure rotational and rovibrational ones. This choice is reasonable since some systematic shifts are of the same order in absolute terms for both types of transitions, and thus the experimental effort to control or determine such shifts appears comparable. 

In view of a realistically only moderate future reduction of theoretical QED uncertainty, there is no advantage in pushing the experimental uncertainty to the levels that ultimately might be possible. We shall therefore assume $u^{\rm proj}(f_i^{\rm exp})=1$\,Hz for all transitions, corresponding to a fractional uncertainty $1\times1 0^{-12}$ for a rotational transition (specifically, transition~1 (\HDplus) in Tab.~\ref{tab:Overview of transitions}) and $\approx5\times10^{-15}$ for vibrational transitions~\footnote{The considered transitions have a hyperfine structure. In experimental works on HD$^+$ so far~\cite{Alighanbari2020,Patra2020,Kortunov2021,Alighanbari2023}, a few hyperfine components were measured, from which a ``spin-averaged'' transition frequency was extracted using the theoretical hyperfine structure. In order to reach the high accuracy levels assumed here, an essential requirement is to measure a sufficient number of hyperfine components, so that the ``spin-averaged'' frequency can be obtained by applying a sum rule and thus will not be limited by the theoretical hyperfine structure uncertainty. The required number of measurements is discussed in~\cite{Schiller2018}. In case of homonuclear MHI in states of zero total nuclear spin, the number of components is reduced to two, strongly simplifying the application of the sum rule. This has recently been demonstrated on H$^+_2$ \cite{Schenkel2024,Alighanbari2024}.}.

\vskip .1in
\noindent\textbf{Case I: Measurements in \HDplus, \H2plus and \D2plus}

\noindent In this section, we revisit the most complete scenario, where future MHI measurements (excluding T-MHI) are combined with existing H/D(1S-2S) data to determine the five FCs
$z_1 = \mu_{pd}/m_e$, $z_2=m_p/m_e$, $z_3 = R_{\infty}$, $z_4 = r_p$, $z_5 = r_d$ ($\mu_{pd}/m_e$, where $\mu_{pd}$ is the reduced mass of proton and deuteron, is considered as an equivalent alternative to $m_d/m_e$). We performed systematic searches to determine the optimal sets of transitions. More precisely, we search for transitions that minimize the following figure of merit (FOM): $\text{FOM1} = \prod_{i=1}^{5} u(z_i)/u_{2022} (z_i)$\,. Here $u(z_i)$ is the uncertainty of the adjusted FC $z_i$ resulting from our model.

One important finding of our previous work~\cite{Schiller2024} was that it is very useful to include measurements of transitions between high-lying vibrational levels, whose sensitivities to the involved nucleus-electron mass ratios is positive rather than negative. On the other hand, measurements on such transitions are expected to be experimentally more difficult, because the molecule must first be prepared in an excited - and in case of heteronuclear MHI, metastable - rovibrational level, before applying the spectroscopy radiation. 

This observation led us to test different scenarios based on experimental considerations: 
\begin{itemize}
\item [I-a)] Avoid high-$v$ transitions altogether: Tab.~\ref{tab:no-T without high-v}-a, 
\item[I-b)] Include only one high-$v$ transition: Tab.~\ref{tab:no-T without high-v}-b, 
\item[I-c)] No restrictions: Tab.~\ref{tab:no-T without high-v}-c, 
\item[I-d)] Include only one high-$v$ transition, and avoid the pure rotational transition in \HDplus: Tab.~\ref{tab:no-T without high-v}-d. 
\end{itemize}

In each table, results obtained from an increasing number of measurements $n$ are shown, starting from $n=3$, the required minimum to determine the five involved FCs with the addition of the H(1S-2S) transition and H-D(1S-2S) isotope shift measurements. This is done by adding, at each line, one more transition to the selection shown in the previous line. In practice, optimization according to FOM1 is performed for the largest sets of transitions (typically $n \geq 7$), and smaller sets are determined by removing 1 transition at each step, choosing the transition that causes the smallest increase of FOM1. Several observations can be made on these results:

\vskip .05in
(i) They show that, by carefully selecting the sets of transitions to be measured, impressive improvements on the mass ratios can be reached (see Tab.~\ref{tab:no-T without high-v}c, and dotted lines in SM, Figs.~\ref{fig:no-T mass ratios},~\ref{fig:no-T FOM 1}). $\mu_{pd}/m_e$ and $m_p/m_e$ are improved by respective factors of up to 180 and 240. The Rydberg constant is improved by a factor of 1.8. Determinations of nuclear radii are not as precise as CODATA 2022, but it should be noted that they are about 5 times more precise than their current determination from electronic H/D  spectroscopy, shown in the last line in Tab.~\ref{tab:no-T without high-v}.

\vskip .05in
(ii) Whereas the above projections are obtained with 9 measurements (last line of Tab.~\ref{tab:no-T without high-v} c), appreciable improvements are already possible with 5 (see e.g.~third line of Tab.~\ref{tab:no-T without high-v} b): a factor of 100 for the mass ratios, and a $R_\infty$ determination that almost matches the current one. Above $n=5$ the improvements yielded by each additional measurement become smaller and smaller, until saturation is reached. Further discussion of Tab.~\ref{tab:no-T without high-v} and a pictorial representation of figures of merit are found in SM, sec.~\ref{Appendix: Graphical display of uncertainties - case I}.

\vskip .05in

\begin{table}[h!]
\footnotesize
    \setlength{\tabcolsep}{2pt}
\begin{tabular}{ccccccccc}
\hline
\hline
 $n$ & \multicolumn{3}{c}{$\text{Transitions}$}  & $u_r(z_1)$ & $u_r(z_2)$ & $u_r\left(R_{\infty }\right)$ & $u(r_p)$ & $u(r_d)$ \\
  & $\text{HD}^+$ & $\text{H}_2^+$ & $\text{D}_2^+$ & $\left(10^{-12}\right)$ & $\left(10^{-12}\right)$ & $\left(10^{-12}\right)$ & $\text{(am)}$ & $\text{(am)}$ \\
\hline
\multicolumn{9}{c}{$\text{Scenario I-a: baseline}$} \\
 3 & \text{\emph{1},3} & \text{} & 13 & 30 & 30 & 11 & 12 & 4.9 \\
 4 & \text{\emph{1},3,4} & \text{} & 13 & 12 & 12 & 1.6 & 1.8 & 0.70 \\
 5 & \text{\emph{1},2,3,4} & \text{} & 13 & 1.1 & 1.1 & 1.2 & 1.4 & 0.57 \\
 6 & \text{\emph{1},2,3,4} & 9 & 13 & 1.1 & 1.1 & 1.2 & 1.4 & 0.55 \\
\hline
\multicolumn{9}{c}{\text{Scenario I-b: includes one ``high-$v$'' transition: \textbf{5} }} \\
 3 & \text{4,\textbf{5}} & \text{} & 13 & 5.1 & 3.4 & 25. & 27. & 11. \\
 4 & \text{\emph{1},4,\textbf{5}} & \text{} & 13 & 0.63 & 0.36 & 4.6 & 5.0 & 2.0 \\
 5 & \text{\emph{1},3,4,\textbf{5}} & \text{} & 13 & 0.16 & 0.17 & 1.2 & 1.5 & 0.57 \\
 6 & \text{\emph{1},3,4,\textbf{5}} & 9 & 13 & 0.16 & 0.092 & 1.2 & 1.4 & 0.55 \\
 7 & \text{\emph{1},2,3,4,\textbf{5}} & 9 & 13 & 0.15 & 0.090 & 1.2 & 1.4 & 0.54 \\
\hline
\multicolumn{9}{c}{\text{Scenario I-c: without any restriction on the choice of transitions}} \\
 3 & \text{\emph{1},\textbf{5}} & 9 & \text{} & 11. & 2.9 & 16. & 17. & 6.9 \\
 4 & \text{\emph{1},2,\textbf{5}} & 9 & \text{} & 2.9 & 2.5 & 1.7 & 1.9 & 0.74 \\
 5 & \text{\emph{1},2,4,\textbf{5}} & 9 & \text{} & 0.16 & 0.32 & 1.3 & 1.5 & 0.58 \\
 6 & \text{\emph{1},2,4,\textbf{5},\textbf{8}} & 9 & \text{} & 0.11 & 0.31 & 0.72 & 0.98 & 0.38 \\
 7 & \text{\emph{1},2,4,\textbf{5},\textbf{8}} & 9 & 13 & 0.11 & 0.076 & 0.69 & 0.92 & 0.37 \\
 8 & \text{\emph{1},2,4,\textbf{5},\textbf{8}} & 9 & \text{13,14} & 0.099 & 0.075 & 0.62 & 0.87 & 0.34 \\
 9 & \text{\emph{1},2,4,\textbf{5},\textbf{6},\textbf{8}} & 9 & \text{13,14} & 0.096 & 0.072 & 0.61 & 0.86 & 0.34 \\
\hline
 \multicolumn{9}{l}{\text{Scenario I-d:}}\\
  \multicolumn{9}{l}{\text{including 1 ``high-$v$'' transition,  excluding the rotational transition \emph{1} }} \\
 3 & \text{4},\textbf{6} & \text{} & 13 & 11. & 9.7 & 29. & 31. & 12. \\
 4 & \text{2,4},\textbf{6} & \text{} & 13 & 0.29 & 0.37 & 14. & 15. & 6.0 \\
 5 & \text{2,4},\textbf{6} & 9 & 13 & 0.17 & 0.28 & 4.7 & 4.9 & 2.0 \\
 6 & \text{2,3,4},\textbf{6} & 9 & 13 & 0.13 & 0.28 & 4.4 & 4.6 & 1.8 \\
\hline
\multicolumn{4}{c}{\text{CODATA 2022}} & 17 & 17 & 1.1 & 0.64 & 0.27 \\
\multicolumn{4}{c}{\text{without $\mu$H, $\mu$D data}} & 17 & 17 & 3.9 & 4.3 & 1.7 \\
 \hline
 \hline
\end{tabular}

    \caption{
    LSA determination of the five quantities $z_1=\mu_{pd}/m_e$, $z_2=m_p/m_e$, $R_\infty$, and proton and deuteron charge radii, by measuring $n$ transitions in \HDplus, \H2plus, and \D2plus at the 1-Hz accuracy level. H(1S-2S) and H-D(1S-2S) measurements are included as input data. Columns 2 -- 4 indicate the chosen transitions, using the numbering defined in Tab.~\ref{tab:Overview of transitions}. Transitions that occur from metastable lower levels in the heteronuclear molecule are indicated in bold. They are ``high-$v$" transitions, as is also number 14. Fractional or absolute uncertainties of the five adjusted FCs are shown in columns 5 -- 9. $u_{\rm QED}/f_i = 8\times 10^{-12}$. We assumed $u_0^{\rm proj} = 1.25 \times 10^{-11}$ (see End Matter and Eq.~\ref{eq:ubethelog} therein for the definition of $u_0$). Transition 1 is the pure rotational transition in \HDplus. The rightmost three entries in the last line are from Tab. XVI of~\cite{Mohr2025}.
     }
     \label{tab:no-T without high-v}
\end{table}

\begin{table*}[htb!]
\footnotesize
     $\begin{array}{ccccccccccccc}
\hline
 n & \multicolumn{5}{c}{\text{Transitions}} & u_r(z_1) & u_r(z_2) & u_r(z_3) & u_r\left(R_{\infty }\right) & u\left(r_p\right) & u\left(r_d\right) & u\left(r_t\right) \\
 \text{} & \text{HD}^+ & \text{H}_2^+ & \text{D}_2^+ & \text{HT}^+ & \text{T}_2^+ & \left(10^{-12}\right) & \left(10^{-12}\right) & \left(10^{-12}\right) & \left(10^{-12}\right) & \text{(am)} & \text{(am)} & \text{(am)} \\
\hline
 7 & 2,4,{\bf 8} & \text{} & 13 & 18  & \text{23,24} & 0.28 & 0.38 & 0.28 & 1.0 & 1.3 & 0.49 & 0.59 \\
 8 & 1,2,4,{\bf 8} & \text{} & 13 & 18  & \text{23,24} & 0.28 & 0.38 & 0.28 & 0.78 & 1.0 & 0.40 & 0.48 \\
 9 & 1,2,4,{\bf 8} & 9 & 13 & 18 &   \text{23,24} & 0.25 & 0.27 & 0.23 & 0.76 & 0.98 & 0.39 & 0.46 \\
 10 & 1,2,3,4,{\bf 8} & 9 & 13 & 18 &   \text{23,24} & 0.25 & 0.26 & 0.23 & 0.75 & 0.97 & 0.39 & 0.46 \\
\hline
 \multicolumn{6}{c}{\text{CODATA 2022}} & 17 & 17 & 38 & 1.1 & 0.64 & 0.27 & 86 \\
\hline
\end{array}
$

     \caption{
     LSA determination of the seven FCs $z_1=m_p/m_e$, $z_2=m_d/m_e$, $z_3=m_t/m_e$, $R_\infty$, $r_p$, $r_d$, $r_t$ by measuring $n$ transitions at the 1-Hz accuracy level. H(1S-2S) and H-D(1S-2S) measurements are included as input data. Columns 2-7 indicate the chosen transitions, using the numbering defined in Tab.~\ref{tab:Overview of transitions}. Fractional or absolute uncertainties of the seven adjusted FCs are shown in columns 8-14. Included is only one ``high-$v$'' transition and three transitions in two -- not three -- T-MHI. $u_{\rm QED}/f_i = 8\times 10^{-12}$, $u_0^{\rm proj} = 1.25 \times 10^{-11}$.
     }
     \label{tab:T 3 TMI 1 high-v}
\end{table*}

\noindent\textbf{Case II: Measurements in all MHI isotopologues}

\noindent We now consider sets of measurements in all isotopologues, including the T-MHI, so that in addition to the previous FCs, also  the triton-electron mass ratio $z_3 = m_t/m_e$, and the the triton charge radius $z_7 = r_t$ can be obtained. Thus, in total, 7 FCs are determined from the LSA. We focus on the tritium-related constants $m_t/m_e$ and $r_t$, without losing sight of the other FCs.

We used two different FOM for systematic searches of optimal sets of transitions. The first one is the generalization of FOM1, i.e.~the product of uncertainties of the 7 FCs, normalized to their reference uncertainty values $u_{2022}$. For $r_t$, which is not included in the CODATA adjustment, we use the value of ref.~\cite{Amroun1994}. The second one is a restriction to tritium-related constants: $\text{FOM1}'=[u(m_t/m_e)/u_{\rm 2022}(m_t/m_e)]\times [u(r_t)/u_{\rm Amroun1994}(r_t)]$. The number of measurements is varied between $n=5$ (the required minimum to determine 7 FCs, with the inclusion of the H(1S-2S) and H-D(1S-2S) isotope shift measurements) and $n=12$, for which the attained precision was observed to saturate in all cases. Similarly to the previous section, taking experimental considerations into account we studied several scenarios.
A detailed discussion is given in SM, sec.~\ref{Appendix: Case II}. 

We single out one scenario that is a compromise between experimental effort and outcome and present it in Tab.~\ref{tab:T 3 TMI 1 high-v}. We find a 150-fold and nearly 200-fold reduction of the uncertainties of $m_t/m_e$  and $r_t$, respectively. Only 3 transitions in two T-MHI need to be measured, and only one ``high-$v$" transition in HD$^+$. In this scenario, the uncertainties of $m_p/m_e$ and $m_d/m_e$ are a factor $\simeq3.5$ higher compared to case I-c (Tab.~\ref{tab:no-T without high-v}-c), but the uncertainties of $r_p$ and $r_d$ are comparable.

\noindent\textbf{Discussion}

\noindent\emph{Dependence on numerical and QED uncertainties.} 

Computations were performed assuming larger numerical uncertainties, $u_0^{\rm proj} = 10^{-10}$ and $u_0^{\rm today} = 10^{-9}$ (SM, sec.~\ref{Appendix: example with larger value of Bethe log uncertainty}). With the latter value, which corresponds to the current state of the art, the FC determinations are worse by factors between 6 and 40. We also investigated the impact of reducing the QED uncertainty by a substantial factor 8 (SM, sec.~\ref{Appendix: examples with smaller value of the QED uncertainty}) under the assumption of low numerical uncertainties ($u_0^{\rm proj} = 1.25 \times 10^{-11}$). We find that this leads to only a small improvement of the charge radii, and not of the mass ratios or Rydberg constant. These results imply that the priority for theoretical efforts should be the (more easily tractable) reduction of the numerical uncertainty of the Bethe logarithm values rather than of the QED uncertainty. Finally, we also investigated the impact of possible deviations from the assumption of perfect correlation between uncertainties (see End Matter) stemming from a given QED term (SM, sec.~\ref{Appendix - Imperfect correlations}). We find that the precision of FC determinations is quite sensitive to such deviations, but that this sensitivity could be greatly reduced by calculating QED terms that are easier to calculate than those which presently limit the theoretical uncertainty.

\noindent\emph{The triton charge radius.}

In order to perform a precision test of chiral effective field theory, $r_t$ should be measured with uncertainty $u_{\rm goal}(r_t)=0.9\,$am or less (SM, sec.~\ref{Appendix - triton charge radius}). The present analysis demonstrates that this is possible.

\noindent\emph{The triton mass.} 

$m_t$ has been measured for several decades, one motivation being to support the work towards improved limits on the electron-weighted neutrino mass (see SM, sec.~\ref{Appendix - triton mass}). If a value of $m_t/m_e$ would eventually be obtained from T-MHI spectroscopy, it could be combined with available or future results from \H2plus\ and \HDplus\ spectroscopy to yield $(m_p+m_d)/m_t$. \
For example, in the scenario described in the first line of Tab.~\ref{tab:T 3 TMI 1 high-v} (with 7 transitions), it could be determined with a fractional uncertainty of $1.3 \times 10^{-13}$. The precision would be further improved to $9.7 \times 10^{-14}$ in the more experimentally challenging scenario given in the last line of Tab.~\ref{tab:T best B} with 12 transitions (SM, sec.~\ref{Appendix: Case II}). After correcting for the binding energy of \HDplus~\cite{Korobov2017a}, $m({\rm HD}^+)/m({\rm T}^+)$ would then be obtained with essentially the same accuracy and may be compared with the direct mass ratio result of ref.~\cite{MedinaRestrepo2023}, that has uncertainty close to $1\times10^{-11}$. This comparison will provide an important cross-check between laser spectroscopy and mass spectrometry. 

\noindent\emph{Spectroscopy of tritium-containing MHIs.}

We expect that trapping, cooling and spectroscopy of \HTplus, \DTplus, and \T2plus can be implemented along the same lines as has already been done for \HDplus and \H2plus. Overall, the spectroscopy will qualitatively be the same, only the transition wavelengths will change, which may require technical adaptations on the laser systems. Additional comments are given in SM, sec.~\ref{Appendix - T-MHI spectroscopy}. Although the T-MHI are radioactive, a property that will make work on them challenging, it is feasible, given that only very small amounts of neutral molecular gas  are required for experimentation on the ions. Clearly, spectroscopy of T-MHI can be implemented and has some interesting advantages as compared to atomic T  spectroscopy (SM, sec.~\ref{Appendix - Atomic T spectroscopy}). On the experimental side these are: use of established techniques, better control over systematics, and less effort in working with the radioactive tritium gas, since only very modest amounts are needed. Importantly, T-MHI spectroscopy also allows for a determination of the triton mass with high accuracy, which is not feasible via spectroscopy of T.

\noindent\emph{Test of QED.} 

We already pointed out~\cite{Alighanbari2024} that a strongly improved value of $m_p/m_e$ will enable testing the QED calculations of the g-factors of the electron bound in a hydrogen-like ion. At present, such calculations are used in the opposite way:  assumed to be accurate within their estimated theory uncertainty and combined with experimental g-factors and appropriate nuclear mass ratios, they yield $m_p/m_e$~\cite{Koehler2015}.

\noindent\emph{BSM Physics.} 

In addition to advancing the determination of FCs, the data of the MHI isotopologues will also allow improved searches for BSM physics in the electron and baryon sector. The formalism developed in ref.~\cite{Delaunay2023} can be straightforwardly extended. Already, the present analysis shows that the proton and deuteron charge radii may be determined using purely ``electronic" H/D and MHI spectroscopy with similar uncertainty as from muonic hydrogen spectroscopy. Therefore, it will be possible to improve constraints for the strength of exotic interactions between leptons and baryons that are different for first and second-generation leptons. This implies a lepton universality test. Future modeling work should address to what extent BSM tests for specific interactions would be improved. 

\noindent\textbf{Conclusion}
\newline
We have shown that pushing the accuracy of MHI spectroscopy and extending it to two tritium-containing isotopologues promises notable improvements in fundamental constants and fundamental physics tests. Such an endeavor is therefore well worth undertaking. No comparable improvements of the named FCs are on the horizon for any other system.

\begin{acknowledgments}
We thank A.A.\,Filin for helpful discussions. This work has received funding from the European Research Council (ERC) under the European Union\textquoteright s Horizon 2020 research and innovation programme (grant agreement No.~786306, ``PREMOL'') (S.S.). This work was also part of 23FUN04 COMOMET that has received funding from the European Partnership on Metrology, co-financed by the European Union's Horizon Europe Research and Innovation Programme and from by the Participating States. Funder ID: 10.13039/100019599  (J.-Ph.K, S.S.).
\end{acknowledgments}

\bibliographystyle{elsarticle-num}
\bibliography{Bibliography-database_v5, JPKbibliography}

%%%%%%%%%%%%%%%%%%%%%%%%%%%%%%%%%%%%%%%%%%%%%%%%%%%%%%%%%%%%%%%%%%%%
% END MATTER

\section*{End Matter}

\noindent\textbf{Sensitivity coefficients}

\noindent Sensitivities to FCs are obtained, with sufficient accuracy, in a nonrelativistic approximation, in the form of operator expectation values over nonrelativistic wave functions of ro-vibrational states (see~\cite{Karr2023} for detailed formulas). The three-body wavefunctions are obtained by the variational method presented in~\cite{Korobov2000}. Precise values of nonrelativistic energies of a large number of ro-vibrational levels of HD$^+$, H$_2^+$, and D$_2^+$ (see e.g.~\cite{Schiller2005,Karr2006,Korobov2017b}) as well as relevant operator expectation values~\cite{Aznabayev2019}, are available. HT$^+$ nonrelativistic energies were published in~\cite{Bekbaev2011}. Calculations in the remaining isotopologues DT$^+$ and T$_2^+$ were performed for the present work (DT$^+$ energy levels were also computed in~\cite{Tian2012}). 

\vskip .1in
\noindent\textbf{Experimental uncertainties and their covariances}

\noindent As stated in the main text, we assume the same experimental uncertainty $u^{\text{proj}}(f_i^{\text{exp}}) = 1$~Hz for all transitions. Furthermore, all experimental data are assumed to be uncorrelated. The covariance of experimental transition frequencies $f_i^{\text{exp}}$ and $f_j^{\text{exp}}$ is then
\begin{equation}
V_{ij}^{\rm exp} = \delta_{ij} \left[ u^{\text{proj}}(f_i^{\text{exp}}) \right]^2 \,,
\end{equation}
where $\delta_{ij}$ is a Kronecker delta symbol.

\vskip .1in
\noindent\textbf{Theoretical uncertainties and their covariances}

 \noindent Similarly to our previous work~\cite{Schiller2024}, theoretical uncertainties and correlations between them are estimated in accordance with the results of ref.~\cite{Korobov2021}; details can be found in the Supplemental Material of~\cite{Delaunay2023}. In the present work, we improve our treatment by taking into account also estimated numerical uncertainties of calculated QED correction terms. So far, these uncertainties had been neglected by us, being much smaller than those due to yet uncalculated terms~\cite{Korobov2021}. Nevertheless, they become important in the context of high-precision determination of FCs from ensembles of measurements. The total theoretical uncertainty of a transition frequency $f_i$ can be written as
 \begin{equation}
\left[u(\delta f_i^{\rm theor})\right]^2=\sum_k \left( u_{{\rm QED},k} (\delta f_i^{\rm theor}) \right)^2 + \left( u_{\rm num} (\delta f_i^{\rm theor}) \right)^2 \,,
 \end{equation}
where $u_{{\rm QED},k} (\delta f_i^{\rm theor})$ are uncertainties associated with estimates of uncalculated QED terms (see~\cite{Korobov2021} for details), and $u_{\rm num} (\delta f_i^{\rm theor})$ the numerical uncertainty.  As mentioned, neglecting the numerical uncertainties, $u(\delta f_i^{\rm theor})\simeq f_i\times8\times10^{-12}$ for the transitions considered in this work. The covariance of theoretical uncertainties of transition frequencies $f_i$ and $f_j$ is given by
\begin{align}
V_{ij}^{\rm theor} &= \text{Cov}(\delta f_i^{\rm theor},\delta f_j^{\rm theor}) \nonumber \\
&= \sum_k r_{k,ij} u_{{\rm QED},k} (\delta f_i^{\rm theor}) u_{{\rm QED},k} (\delta f_j^{\rm theor}) \nonumber \\
&+ \delta_{ij} u_{\rm num} (\delta f_i^{\rm theor}) u_{\rm num} (\delta f_j^{\rm theor}) \label{eq:cov matrix}
\end{align}
\noindent where $r_{k,ij}$ is a correlation coefficient. The complete covariance matrix of the input data for the LSA is $V = V^{\rm exp} \oplus V^{\rm theor}$, meaning that $V$ is a block diagonal matrix constructed from the $V^{\rm exp}$ and $V^{\rm theor}$ matrices. In the main part of this paper, we assume QED uncertainties to be perfectly correlated, i.e. $r_{k,ij}=1$, while numerical uncertainties are uncorrelated and do not contribute in the off-diagonal covariances (note the Kronecker symbol in the last line), lowering the overall correlation. 

As discussed in~\cite{Schiller2024}, it is the strong correlation between different ro-vibrational transitions that allows unknown QED contributions to be determined from experimental data, and subsequently the FCs to be determined with high accuracies. This has two important implications. The first one is that the above perfect-correlation hypothesis has strong impact on the outcome of our analysis, and should therefore be carefully examined. While this hypothesis is a well motivated approximation (see discussion in SM, sec.~\ref{Appendix - Imperfect correlations}), it is not exact. We therefore analyzed the impact of possible deviations from this hypothesis in the SM (sec.~\ref{Appendix - Imperfect correlations}). The second implication is that ignoring numerical uncertainties may lead to overestimating the correlations and thus to over-optimistic projections, especially in the limit of high experimental accuracies.

As a consequence, it is important to make a realistic estimate of numerical uncertainties. Most relativistic and QED correction terms are given by operator expectation values that can be calculated with high accuracies~\cite{Aznabayev2019}, which can be improved further by several orders of magnitude at a moderate computational cost. The only exceptions are the quantities of the ``Bethe logarithm'' type, which pose more serious numerical difficulties. Among them, the largest contribution by far is from the nonrelativistic Bethe logarithm $\beta_{v,N}$~\cite{Korobov2012} (where $v,N$ denotes the rovibrational level), at the $m_e\alpha^5$ order, whereas the next ones arise at the $m_e\alpha^7$ order~\cite{Korobov2013b}. It is therefore reasonable to neglect sources of numerical uncertainty other than the nonrelativistic Bethe logarithm. The numerical uncertainty of a given rovibrational transition frequency is then 
\begin{equation}
u_{\rm num} (\delta f_i^{\rm theor}) = (2 R_{\infty}c) \frac{4 \alpha^3}{3} \left( u(\beta_{v,N})^2 + u(\beta_{v',N'})^2 \right)^{1/2} \,, \label{eq:unum}
\end{equation}
where $(v,N)$ and $(v',N')$ are the rovibrational states connected by the transition under consideration. 

While it is certainly possible to improve the accuracy of existing Bethe logarithm calculations~\cite{Korobov2012}, it remains true than with a given amount of computational resources the numerical uncertainty will be an increasing function of vibrational and (to a lesser extent) rotational excitation. The dependence of numerical uncertainties on the ro-vibrational state $(v,N)$ should be at least approximately accounted for in order to perform realistic projections of FC determinations. Analysis of the results from Tab. IV and V of ref.~\cite{Korobov2012}, where comparable resources were devoted to each ro-vibrational state, shows that the uncertainties as estimated in that work are reasonably described by the following simple formula:
\begin{equation}
u(\beta_{v,N}) = u_0 (v+1)^{3/2} (N+1)^{1/2} , \label{eq:ubethelog}
\end{equation}
with $u_0$ in the few 10$^{-9}$ range. In this work, we add this contribution to the other sources of uncertainty discussed in~\cite{Korobov2021}, and assume that uncertainties affecting different ro-vibrational levels are uncorrelated to each other. Furthermore, we assume a significantly lower value of $u_0$ with respect to what can be estimated by matching with the results of~\cite{Korobov2012}, reflecting the fact that Bethe logarithm calculations can be improved in the future with dedicated efforts. We consider the value $u_0^{\rm proj} = 1.25 \times 10^{-11}$, which represents an improvement by a factor of order 100 for the ground ro-vibrational state. Larger values of $u_0$, including the present state-of-the-art, $u_0^{\text{today}}=10^{-9}$, are considered in SM, Sec.~\ref{Appendix: example with larger value of Bethe log uncertainty}.

\vfill\eject
\clearpage
\newpage

%%%%%%%%%%%%%%%%%%%%%%%%%%%%%%%%%%%%%%%%%%%%%%%%%%%%%%%%%%%%%%%%%%%
% SUPPLEMENTAL MATERIAL

\onecolumngrid
\begin{center}
{\Large \textbf{Supplemental Material}}
\end{center}
\twocolumngrid

\vskip .1in
\renewcommand\appendixname{}
\appendix

\section{The triton charge radius}
\label{Appendix - triton charge radius}

We explain the main motivation for extending transition frequency measurements to T-MHI, i.e. those containing one or two tritons. 

The mass of the triton relative to the electron mass, $m_t/m_e$ and the triton radius $r_t$ are the two quantities that may be extracted from appropriate measurements on T-MHI. In this section we discuss why obtaining an accurate value of the triton radius is of substantial interest.

A hot topic in nuclear physics is to push the theory accuracy further, with the goal of explaining accurately nuclear properties, for example the radii of nuclei with a moderate number of nucleons. The theory must first develop an accurate description of the few-nucleon interactions. Experiments can provide input data for the development of the theory.

The most fundamental interactions are the two-nucleon ones, proton-proton and proton-neutron. They can be studied in scattering (collision) experiments, and the latter one also in the two-nucleon bound system deuteron, for example through its charge radius and electric quadrupole moment. Similarly, the interactions of more nucleons may be studied in successively larger nuclei or in more complicated scattering experiments. The simplest case for the latter is neutron - deuteron scattering, yielding information about so-called 3N interactions. Further interactions of interest are photon-nucleon interactions, and partial information comes form the form factors of proton and neutron.

Chiral effective field theory ($\chi$EFT) is an ab initio nuclear theory where the nucleon-nucleon interactions are systematically treated. Its accuracy has been steadily improved over the course of many years and it can now precisely analyze scattering reactions and nuclear structure~\cite{Epelbaum2009,Epelbaum2012,Epelbaum2020,Machleidt2011}. Recently, precision $\chi$EFT has been expanded beyond the two-nucleon case (the deuteron), and predictions for properties of the triton, and of the $^3$He and $^4$He nuclei have been worked out. These predictions include a quantity related to the charge radius of a nucleus, the so-called structure radius $r_{\rm str}$ (see below). 

Charge radii can be measured experimentally by electron scattering or - indirectly - through tiny shifts that they cause to energy levels of atomic and molecular levels. QED theory predictions for the latter systems are necessary in order to extract the charge radii from spectroscopic data. Thus, if both accurate experimental atomic/molecular data, accurate QED predictions for them, and accurate $\chi$EFT predictions are available, there is an excellent opportunity to perform tests of $\chi$EFT theory predictions. Successful tests have several implications~\cite{Filin2023}:

 - extraction of the proton and the neutron charge radius from few-nucleon data

- support the development of predictions for the charge radii of medium-mass and heavy nuclei

- search for beyond-standard-model physics in the nuclear realm

$\chi$EFT does not compute the charge radius $r$ per se, but the so-called structure radius $r_{\rm str}$. For a nucleus $X$ the two are related by
\begin{equation}
    r^2(X)=r_{\rm str}^2(X)+\left(r_p^2+\frac{3}{4\,m_p^2}\right)+\frac{A-Z}{Z}\,r_n^2\ .
\label{eq:r^2}
\end{equation}
Here $A$, $Z$ are the mass number and proton number of the nucleus in question, $r_p$, $r_n$ are the charge radii of proton and neutron. Units are such that $1/m_p$ is the reduced Compton wavelength of the proton. Currently, $\chi$EFT predicts
\begin{equation}
\begin{split}
        r_{\rm str}(d)&=1.9729(15)\,{\rm fm}\ ,\\
    r_{\rm str}(^4{\rm He})&\simeq1.4758(32)\,{\rm fm}\ ,\\
    r_{\rm str}({\rm isoscalar\, ^3H})&\simeq1.7312(21)\,{\rm fm}\ .\\
\end{split}
\end{equation}
The first value is from ref.\,\cite{Filin2021}, the last two values are preliminary \cite{Filin2024privcomm}. The isoscalar structure radius is a combination of structure radius of the triton and helium-3 nucleus:
\begin{equation}
    r_{\rm str}^2({\rm isoscalar\, ^3H})=\frac{1}{3}(r_{\rm str}^2({\rm^3H})+2\,r_{\rm str}^2({\rm^3He}))\ .
    \label{eq:r_str isoscalar}
\end{equation}

The experimental data available for testing the structure radius predictions are charge radii. This data comes from electron scattering (a technique that has led to the determination of all charge radii, including that of the neutron), from atomic H, D, $^3$He, $^4$He spectroscopy and from muonic H, D, $^3$He, $^4$He spectroscopy. Because of the form of eq.\,(\ref{eq:r^2}), at least three experimental data must be combined to test the $\chi$EFT prediction $r_{\rm str}$. 
Based on eq.(\ref{eq:r^2}), $\chi$EFT predictions were tested concerning:

(A) $^4$He charge radius: the $\chi$EFT prediction (2N-EFT \& 4N-EFT, with experimental $r_p$ and 1s-2s H/D isotope shift data as input) is in agreement with both the electron scattering result (having similar accuracy) and with the (more accurate) muonic $^4$He measurement;

(B) neutron charge radius: the prediction from 2N-EFT plus H/D isotope shift data agrees with determination from low-energy neutron attenuation by electron-rich targets.

(C) proton charge radius: the value deduced from 2N-EFT \& 4N-EFT and experimental data on 1S-2S H-D isotope shift and on the $^4$He charge radius is in agreement with (more accurate) recent determinations from electronic and muonic spectroscopy of H~\cite{Maisenbacher2024}.

(D) $\chi$EFT can also be tested concerning the 3N-prediction. As eq.\,(\ref{eq:r_str isoscalar}) indicates, one necessarily requires experimental data on both $^3$He and $^3$H, the triton. Furthermore, from eq.\,(\ref{eq:r^2}) one obtains
\begin{equation}
      r_{\rm str}^2({\rm isoscalar\, ^3H})-r_{\rm str}^2(d)=
      \frac{1}{3} r_t^2+\frac{2}{3}r^2({\rm^3He})-r_d^2\ .
    \label{eq:relationship to test}
\end{equation}
On the left hand side are theoretically predicted quantities, while on the right hand side are quantities to be measured experimentally. We emphasize that an agreement between the two sides of the equation is nontrivial, as it does not follow from already performed tests.

The total theory uncertainty is $u({\rm l.h.s})=0.0090\,{\rm fm}^2$, where the two individual theory uncertainties may be assumed uncorrelated, since they result mostly from truncation errors. 

With the values $r_d=2.127\,78(27)\,$fm, the recently obtained $r({\rm ^3He})=1.9701(9)$\,fm \cite{CREMA2023helion} and a much older triton charge radius from electron scattering, $r_t =1.755(86)$\,fm \cite{Amroun1994},  eq.\,(\ref{eq:relationship to test}) is satisfied within the uncertainties. On the r.h.s., the uncertainties of the three terms are $0.10\,{\rm fm}^2$, $0.0024\,{\rm fm}^2$, and $0.0011\,{\rm fm}^2$, respectively. 

Thus, because the triton charge radius is much less accurately known than for $d$ and $^3$He, it dominates the total uncertainty of the r.h.s., which is 11 times larger than the uncertainty of the prediction (the l.h.s.). This situation is the motivation for a high-accuracy measurement of the triton radius. Without it, an important prediction of $\chi$EFT cannot be tested at a relevant level.

The current uncertainty of the charge radius of $^3$He sets the accuracy goal of a future determination of $r_t$: its uncertainty should contribute less than that of $^3$He, i.e. $u(r_t) \ll 2 u(r(^3{\rm He})) = 0.0018\,{\rm fm}$. Setting the goal $u_{\rm goal}(r_t)=0.0009$\,fm would imply an approximately one-hundred-fold reduction compared to the current uncertainty. An intermediate goal is a total experimental uncertainty moderately smaller than the current theoretical one, implying $u_{\rm goal}(r_t)\simeq0.003\,{\rm fm}$, 30 times smaller than available today.

%%%%%%%%%%%%%%%%%%%%%%%%%%%%%%%%%
\section{The triton mass $m_t$}
\label{Appendix - triton mass}

The triton mass has been measured for several decades, one motivation being to support the work towards improved limits on the electron-weighted neutrino mass. In the experiment KATRIN, this is done through the study of tritium $\beta$-decay~\cite{KATRINCollaboration2025short}, where the electron spectrum is analyzed near its end point. This is where the energy difference between the masses of T and $^3$He (the Q-value) plays a crucial role. The Q-value measured by KATRIN is $18\,591.49(50)\,$eV.

Independent determinations of the Q-value help verify the results of these experiments and provide an improved check on their systematics. The most powerful approach uses Penning trap mass spectrometry to determine the mass difference of tritium and $^3$He and derives the Q-value. By measuring the cyclotron frequency ratios of T$^+$ and $^3$He$^+$, as well as of the two pairs HD$^+$/$^3$He$^+$ and HD$^+$/T$^+$, a Q-value 18592.071(22)\,eV was obtained by Medina Restrepo and Myers~\cite{MedinaRestrepo2023}. It is in agreement with KATRIN~\cite{KATRINCollaboration2025short}, but 22~times more precise.

%%%%%%%%%%%%%%%%%%%%%%%%%%%%%%%%%
\section{List of remaining transitions}

In Tab.~\ref{tab:Overview of remaining transitions}, we give the properties of all the transitions that were considered in our analysis, but are not included in the scenarios described in the main text (and thus not listed in Tab.~\ref{tab:Overview of transitions}). Several of these transitions appear in the scenarios presented in secs.~\ref{Appendix: Case II}-\ref{Appendix - Imperfect correlations}.

\begin{table*}[htb!]
\begin{centering}
\[
\begin{array}{cccc|ccccccccc}
 i & \text{Molecule} & \text{Transition} & f_i & \multicolumn{7}{c}{\text{Normalized sensitivity to the constant $z$ (kHz)}}  \\
 \text{} & \text{} & (v,N)\rightarrow (v',N') & \text{(THz)} & m_p/m_e & m_d/m_e & m_t/m_e & c\,R_{\infty } & r_p & r_d & r_t \\
\hline
 7 & \text{HD}^+ & (9,1\text{)$\to $(}13,0) & 118 & -0.0096 & -0.0048 & 0 & 0.13 & -0.045 & -0.046 & 0 \\
 10 & \text{H}_2^+ & (11,0\text{)$\to $(}13,2) & 53 & 0.32 & 0 & 0 & 0.058 & -0.040 & 0 & 0 \\
 11 & \text{H}_2^+ & (12,0\text{)$\to $(}14,2) & 46 & 0.48 & 0 & 0 & 0.050 & -0.033 & 0 & 0 \\
 12 & \text{H}_2^+ & (9,0\text{)$\to $(}11,2) & 68 & 0.033 & 0 & 0 & 0.074 & -0.052 & 0 & 0 \\
 15 & \text{D}_2^+ & (16,0\text{)$\to $(}18,2) & 38 & 0 & 0.21 & 0 & 0.042 & 0 & -0.030 & 0 \\
 16 & \text{D}_2^+ & (17,0\text{)$\to $(}19,2) & 34 & 0 & 0.28 & 0 & 0.038 & 0 & -0.027 & 0 \\
 17 & \text{HT}^+ & (0,0\text{)$\to $(}0,1) & 1.2 & -0.015 & 0 & -0.011 & 0.0013 & -0.00087 & 0 & -0.24 \\
 19 & \text{HT}^+ & (0,0\text{)$\to $(}5,1) & 247 & -1.4 & 0 & -1.0 & 0.27 & -0.11 & 0 & -30. \\
 20 & \text{HT}^+ & (11,1\text{)$\to $(}15,0) & 100 & 0.15 & 0 & 0.11 & 0.11 & -0.039 & 0 & -9.8 \\
 21 & \text{DT}^+ & (0,0\text{)$\to $(}0,1) & 0.7 & 0 & -0.0075 & -0.011 & 0.00081 & 0 & -0.00059 & -0.15 \\
 22 & \text{DT}^+ & (0,0\text{)$\to $(}5,1) & 201 & 0 & -0.92 & -1.4 & 0.22 & 0 & -0.093 & -24. \\
\hline
\end{array}
\]
\par
\end{centering}
\caption{\label{tab:Overview of remaining transitions}
\footnotesize 
Sensitivities of various transitions of five MHI to the most relevant fundamental constants $z_j$. This table complements Tab.~\ref{tab:Overview of transitions}. Note the change in sign of the mass sensitivities for transitions 10, 11, 20 compared to the transitions 1 -- 4 in Tab.~\ref{tab:Overview of transitions},  and their near-vanishing values for transitions 7 and 12.
}
\end{table*}

%%%%%%%%%%%%%%%%%%%%%%%%%%%%%%%%%%%%%%%
\section{Case I (no tritium-containing MHI)}
\label{Appendix: Graphical display of uncertainties - case I}

\begin{figure}[h!]
\begin{centering}
\includegraphics[width=1\columnwidth]{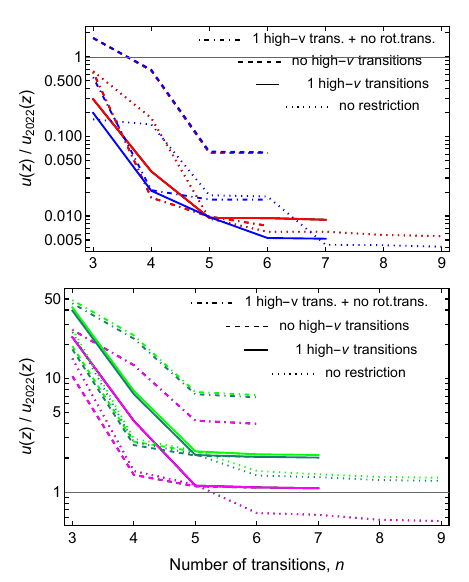}
\par
\end{centering}
\caption{
Upper plot: uncertainties of the $\mu_{pd}/m_e$ (red) and $m_p/m_e$ (blue) mass ratios, normalized to their CODATA 2022 values, as a function of the number of MHI transition measurements. 
Bottom plot: normalized uncertainties of the Rydberg constant (magenta), $r_p$ (green) and $r_d$ (dark green).
\label{fig:no-T mass ratios}}
\end{figure}

\begin{figure}[h!]
\begin{centering}
\includegraphics[width=1\columnwidth]{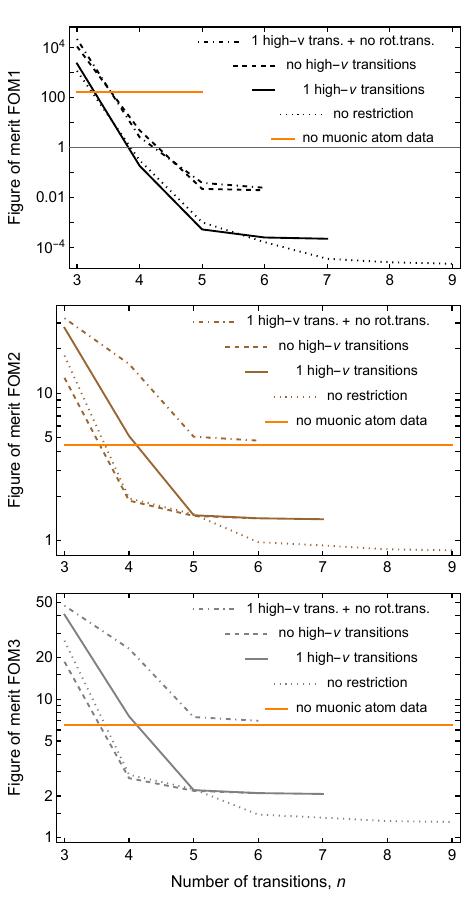}
\par
\end{centering}
\caption{
Upper plot: the figure of merit FOM1, defined as the product of uncertainties of all five FCs normalized to their CODATA 2022 values.
Middle plot: the figure of merit FOM2, defined as the root-mean-square of the normalized uncertainties of all five FCs.
Bottom plot: the figure of merit FOM3, defined as the root-mean-square of the normalized uncertainties of the proton and deuteron charge radii.
The orange lines refer to the CODATA 2022 values obtained excluding muonic atom data (see Tab. XVI of \cite{Mohr2025}).
\label{fig:no-T FOM 1}
}
\end{figure}

Figs.~\ref{fig:no-T mass ratios} and \ref{fig:no-T FOM 1} are graphical displays of FC uncertainties and of three different figures of merit, as obtained in the four scenarios of case I presented in the main text. Continuing the discussion of these scenarios, we remark the following.

(iii) Comparison between Tab.~\ref{tab:no-T without high-v}a and \ref{tab:no-T without high-v}b, and between dashed and full lines in Fig.~\ref{fig:no-T mass ratios}~(upper), confirms the importance of including (at least) one transition between high-lying vibrational levels in order to optimize the determination of the mass ratios. Comparing the results shown in the last line of Tab.~\ref{tab:no-T without high-v}a with those in the fourth line of Tab.~\ref{tab:no-T without high-v}b, one can see that a single replacement of a low-$v$ transition (2) by a high-$v$ one (5) allows improving both mass ratios by respective factors of 7 and 12. It does however not improve the determination of the Rydberg constant and nuclear charge radii, compare dashed and full lines in Fig.~\ref{fig:no-T mass ratios}~(lower). It is also worth noting that the results obtained with one high-$v$ transition are already close to the optimal ones given in Tab.~\ref{tab:no-T without high-v}c: one or two additional transitions of this kind yield further improvement factors between 1.25 and 2, depending on the FC under consideration.

\vskip .05in
(iv) Comparison between  Tab.~\ref{tab:no-T without high-v}b and  \ref{tab:no-T without high-v}d, and between full and dash-dotted lines in Fig.~\ref{fig:no-T mass ratios}~(upper, lower), shows that the pure rotational transition in HD$^+$ plays an important role in improving the determination of $R_\infty$ and $m_p/m_e$.

\clearpage
\section{Case II: sets of transitions including T-MHI}
\label{Appendix: Case II}

\begin{figure}[h]
\begin{centering}
\includegraphics[width=1\columnwidth]{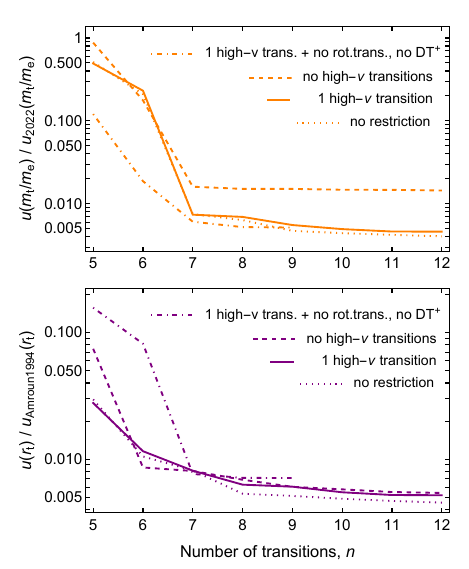}

\par
\end{centering}
\caption{LSA with T-MHI transitions.\\ Upper plot: uncertainty of the $m_t/m_e$ (blue) mass ratio, normalized to its CODATA 2022 value, as a function of the number of MHI transition measurements.
Lower plot: uncertainty of the triton charge radius $r_t$, normalized to its value from Ref.~\cite{Amroun1994}.\\
In both plots, dashed (full) lines correspond to results obtained with zero (one) high-$v$ transition. Dash-dotted lines are obtained with one high-$v$ transition but exclude pure rotational transitions as well as transitions in \DTplus. Finally, dotted lines are obtained without imposing any restrictions. 
\label{fig:T triton-electron mass ratio}}
\end{figure}

\begin{figure}[h]
\begin{centering}
\includegraphics[width=1\columnwidth]{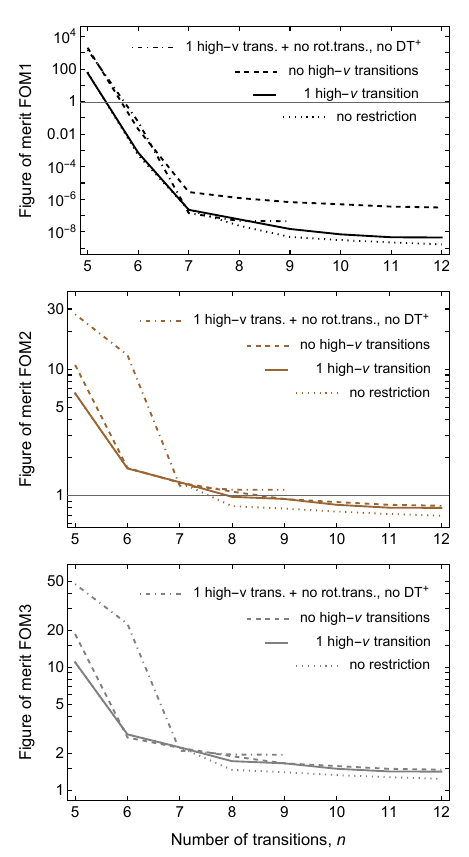}
\par
\end{centering}
\caption{LSAs with T-MHI transitions.\\ Upper plot: the figure of merit FOM1, defined as the product of uncertainties of all seven FCs normalized to their reference values (CODATA 2022 and~\cite{Amroun1994}).\\
Middle plot: the figure of merit FOM2, defined as the root-mean-square of the normalized uncertainties of all seven FCs.\\
Bottom plot: the figure of merit FOM3, defined as the root-mean-square of the normalized uncertainties of the proton and deuteron charge radii.
\label{fig:T FOM}}
\end{figure}

Similar to case I, we studied the following scenarios: 

(i) Measurements of low-$v$ transitions only (Tab.~\ref{tab:T no high-v} and dashed lines in Figs.~\ref{fig:T triton-electron mass ratio}, \ref{fig:T FOM}), 

(ii) sets of measurements including a single high-$v$ transition (Tab.~\ref{tab:T 1 high-v B} and solid lines in Figs.~\ref{fig:T triton-electron mass ratio}, \ref{fig:T FOM}), 

(iii) no restriction on the type of transition (Tab.~\ref{tab:T best B} and dotted lines in Figs.~\ref{fig:T triton-electron mass ratio}, \ref{fig:T FOM}), and 

(iv) selections including a single high-$v$ transition and no pure rotational transition (Tab.~\ref{tab:T easy A} and \ref{tab:T easy B}, dot-dashed lines in Figs.~\ref{fig:T triton-electron mass ratio},~\ref{fig:T FOM}). In this last scenario, we also avoided transitions in \DTplus, which may be more difficult to work with experimentally.

The results of these four scenarios are displayed graphically in Fig.~\ref{fig:T triton-electron mass ratio}, which presents the uncertainties of the triton mass ratio and charge radius, and Fig.~\ref{fig:T FOM}, which shows three figures of merit, FOM1, FOM2, FOM3.

The main findings of the study of case II can be summarized as follows:
\begin{itemize}
    \item The determinations of the triton-electron mass ratio and of the triton charge radius can be improved by very large factors of 250 and 220, respectively. Meanwhile, the precision of the other constants is very similar to what was obtained in the previous section. It is interesting to note that, for example, the selected \HDplus, \H2plus and \D2plus transitions in line 6 of Tab.~\ref{tab:T best B} are exactly the same as in line 6 of Tab.~\ref{tab:no-T without high-v}c; they are complemented by one transition in \HTplus and one in \T2plus. This indicates that experimental campaigns aimed at determination of proton/deuteron observables on the one hand, and triton observables on the other hand, are fully compatible with each other.
    \item In all scenarios, a clear threshold is observed at $n=7$ transitions, where the results are improved by more than one order of magnitude with respect to the $n=6$ case.
    \item Comparison between Tab.~\ref{tab:T no high-v} and \ref{tab:T 1 high-v B} and between dashed and solid lines in the figures, shows again that the inclusion of at least 1 high-$v$ transition is strongly beneficial for the determination of the mass ratios, although the effect is a bit less pronounced than in the previous section, with relative improvement factors of 7, 2.5 and 3.1. Here again, it only has very small impact on the determination of the Rydberg constant and nuclear radii.
    \item The results obtained with 1 high $v$-transition are quite close to the optimal ones; for example, the triton charge radius can be obtained with 0.43~am uncertainty (see Tab.~\ref{tab:T 1 high-v B}) which is very close to the optimal value of 0.39~am (Tab.~\ref{tab:T best B}). It is thus not essential to perform multiple measurements on these experimentally more difficult transitions, at least in a first stage.
    \item Comparison between Tab.~\ref{tab:T 1 high-v B}) and \ref{tab:T easy A} shows that the avoidance or pure rotational transitions slightly degrades the precision of all the FCs, typically by a factor on the order of 1.5.
    \item It is interesting to note that a high number of measurements in tritium-containing isotopologues is not required to get precise determinations of $m_t/m_e$ and $r_t$. For example, excellent results are obtained in line 6 of Tab.~\ref{tab:T best B}, where only 2 out of 10 measurements are performed on TMI. On the other hand, this selection of transitions contains 3 high-$v$ transitions, and inspection of Tab.~\ref{tab:T 1 high-v B} suggests that if one wishes to rely on a single high-$v$ transition measurement, it is then necessary to include more TMI transitions. This triggers the question of whether it is possible to restrict simultaneously the number of high-$v$ transitions and TMI transitions in our selection, while still getting close-to-optimal results. The data from Tab.~\ref{tab:T 2 TMI 1 high-v} provide a first answer: with one high-$v$ transition and 2 TMI transitions, the results are noticeably worse than the optimal ones (e.g. by a factor of about 1.7 for $r_t$). However, increasing the number of TMI transitions to 3 allows to recover close-to-optimal results (e.g. $u(r_t) = 0.46$\,am, worse than the overall best result by only about 20 percent), as can be seen in Tab.~\ref{tab:T 3 TMI 1 high-v}. This scenario eases somewhat the experimental difficulties involved by the proposed measurements, which led us to single it out in the main text.
\end{itemize}

\begin{table*}[h!]
\footnotesize
     \[
\begin{array}{cccccccccccccc}
\hline
 n & \multicolumn{6}{c}{\text{Transitions}} & u_r\left(m_p\right/m_e) & u_r\left(m_d\right/m_e) & u_r\left(m_t\right/m_e) & u_r\left(R_{\infty }\right) & u\left(r_p\right) & u\left(r_d\right) & u\left(r_t\right) \\
 \text{} & \text{HD}^+ & \text{H}_2^+ & \text{D}_2^+ & \text{HT}^+ & \text{DT}^+ & \text{T}_2^+ & \left(10^{-12}\right) & \left(10^{-12}\right) & \left(10^{-12}\right) & \left(10^{-12}\right) & \text{(am)} & \text{(am)} & \text{(am)} \\
\hline
 5 & \text{1,3} & \text{} & 13 & 19 & 22 & \text{} & 30. & 29. & 34. & 11. & 12. & 4.9 & 6.4 \\
 6 & \text{1,3,4} & \text{} & 13 & 19 & 22 & \text{} & 12. & 11. & 6.7 & 1.6 & 1.8 & 0.70 & 0.74 \\
 7 & \text{1,3,4} & \text{} & 13 & 19 & 22 & 23 & 0.64 & 0.59 & 0.61 & 1.3 & 1.5 & 0.58 & 0.69 \\
 8 & \text{1,3,4} & \text{} & 13 & \text{17,19} & 22 & 23 & 0.62 & 0.55 & 0.58 & 1.0 & 1.3 & 0.49 & 0.59 \\
 9 & \text{1,3,4} & \text{} & 13 & \text{17,19} & \text{21,22} & 23 & 0.62 & 0.55 & 0.58 & 0.87 & 1.1 & 0.43 & 0.52 \\
 10 & \text{1,3,4} & \text{} & 13 & \text{17,19} & \text{21,22} & \text{23,24} & 0.59 & 0.55 & 0.56 & 0.80 & 1.1 & 0.41 & 0.50 \\
 11 & \text{1,3,4} & 9 & 13 & \text{17,19} & \text{21,22} & \text{23,24} & 0.54 & 0.52 & 0.56 & 0.77 & 0.99 & 0.39 & 0.47 \\
 12 & \text{1,2,3,4} & 9 & 13 & \text{17,19} & \text{21,22} & \text{23,24} & 0.53 & 0.51 & 0.55 & 0.75 & 0.97 & 0.39 & 0.46 \\
\hline
 \multicolumn{7}{c}{\text{CODATA 2022}} & 17 & 17 & 38 & 1.1 & 0.64 & 0.27 & 86 \\
\hline
\end{array}
\]
     \caption{Similar to Tab.~\ref{tab:T 3 TMI 1 high-v}, but without any ``high-$v$'' transition.
     }
     \label{tab:T no high-v}
\end{table*}

\begin{table*}[h!]
\footnotesize
     \[
\begin{array}{cccccccccccccc}
\hline
 n & \multicolumn{6}{c}{\text{Transitions}} & u_r\left(m_p\right/m_e) & u_r\left(m_d\right/m_e) & u_r\left(m_t\right/m_e) & u_r\left(R_{\infty }\right) & u\left(r_p\right) & u\left(r_d\right) & u\left(r_t\right) \\
 \text{} & \text{HD}^+ & \text{H}_2^+ & \text{D}_2^+ & \text{HT}^+ & \text{DT}^+ & \text{T}_2^+ & \left(10^{-12}\right) & \left(10^{-12}\right) & \left(10^{-12}\right) & \left(10^{-12}\right) & \text{(am)} & \text{(am)} & \text{(am)} \\
\hline
 5 & 4,{\bf 6} & \text{} & 13 & 18 & \text{} & 24 & 9.7 & 14. & 2.5 & 29. & 31. & 12. & 13. \\
 6 & 1,4,{\bf 6} & \text{} & 13 & 18 & \text{} & 24 & 4.2 & 4.4 & 1.0 & 3.0 & 3.2 & 1.3 & 2.2 \\
 7 & 1,4,{\bf 6} & \text{} & 13 & 18 & \text{} & \text{23,24} & 0.27 & 0.48 & 0.24 & 1.1 & 1.3 & 0.53 & 0.61 \\
 8 & 1,2,4,{\bf 6} & \text{} & 13 & 18 & \text{} & \text{23,24} & 0.18 & 0.34 & 0.22 & 0.85 & 1.1 & 0.42 & 0.51 \\
 9 &  1,2,4,{\bf 6} & 9 & 13 & 18 & \text{} & \text{23,24} & 0.12 & 0.23 & 0.17 & 0.82 & 1.0 & 0.41 & 0.49 \\
 10 &  1,2,4,{\bf 6} & 9 & 13 & 18 & 21 & \text{23,24} & 0.12 & 0.22 & 0.16 & 0.72 & 0.94 & 0.37 & 0.45 \\
 11 &  1,2,4,{\bf 6} & 9 & 13 & \text{17,18} & 21 & \text{23,24} & 0.12 & 0.21 & 0.16 & 0.67 & 0.91 & 0.36 & 0.43 \\
 12 &  1,2,4,{\bf 6} & 9 & 13 & \text{17,18,19} & 21 & \text{23,24} & 0.12 & 0.21 & 0.15 & 0.67 & 0.91 & 0.36 & 0.43 \\
\hline
 \multicolumn{7}{c}{\text{CODATA 2022}} & 17 & 17 & 38 & 1.1 & 0.64 & 0.27 & 86 \\
\hline
\end{array}
\]
     \caption{Similar to Tab.~\ref{tab:T no high-v}, but including one ``high-$v$'' transition (here, transition 6 in \HDplus). 
     }
     \label{tab:T 1 high-v B}
\end{table*}

\begin{table*}[h]
\footnotesize
     \[
\begin{array}{cccccccccccccc}
\hline
 n & \multicolumn{6}{c}{\text{Transitions}} & u_r\left(m_p\right/m_e) & u_r\left(m_d\right/m_e) & u_r\left(m_t\right/m_e) & u_r\left(R_{\infty }\right) & u\left(r_p\right) & u\left(r_d\right) & u\left(r_t\right) \\
 \text{} & \text{HD}^+ & \text{H}_2^+ & \text{D}_2^+ & \text{HT}^+ & \text{DT}^+ & \text{T}_2^+ & \left(10^{-12}\right) & \left(10^{-12}\right) & \left(10^{-12}\right) & \left(10^{-12}\right) & \text{(am)} & \text{(am)} & \text{(am)} \\
\hline
 5 & 4,{\bf 8} & \text{} & 13 & 18 & \text{} & 23 & 2.9 & 36. & 3.4 & 16. & 17. & 6.9 & 8.8 \\
 6 & 2,4,{\bf 8} & \text{} & 13 & 18 & \text{} & 23 & 2.5 & 4.0 & 3.0 & 1.7 & 1.9 & 0.74 & 0.77 \\
 7 & 2,4,{\bf 5},{\bf 8} & \text{} & 13 & 18 & \text{} & 23 & 0.32 & 0.67 & 0.41 & 1.3 & 1.5 & 0.58 & 0.72 \\
 8 & 2,4,{\bf 5},{\bf 8} & 9 & 13 & 18 & \text{} & 23 & 0.31 & 0.64 & 0.35 & 0.72 & 0.98 & 0.38 & 0.49 \\
 9 & 1,2,4,{\bf 5},{\bf 8} & 9 & 13 & 18 & \text{} & 23 & 0.076 & 0.24 & 0.17 & 0.69 & 0.92 & 0.37 & 0.44 \\
 10 & 1,2,4,{\bf 5},{\bf 8} & 9 & 13,{\bf 14} & 18 & \text{} & 23 & 0.075 & 0.21 & 0.16 & 0.62 & 0.87 & 0.34 & 0.41 \\
 11 & 1,2,4,{\bf 5},{\bf 8} & 9 & 13,{\bf 14} & 18 & 21 & 23 & 0.073 & 0.20 & 0.15 & 0.58 & 0.84 & 0.33 & 0.40 \\
 12 & 1,2,4,{\bf 5},{\bf 8} & 9 & 13,{\bf 14} & \text{17,18} & 21 & 23 & 0.071 & 0.19 & 0.15 & 0.55 & 0.82 & 0.32 & 0.39 \\
\hline
 \multicolumn{7}{c}{\text{CODATA 2022}} & 17 & 17 & 38 & 1.1 & 0.64 & 0.27 & 86 \\
\hline
\end{array}
\]
     \caption{Similar to Tab.~\ref{tab:T no high-v}, but without any restriction on the choice of transitions. 
     }
     \label{tab:T best B}
\end{table*}

\begin{table*}[h]
\footnotesize
     \[
\begin{array}{cccccccccccccc}
\hline
 n & \multicolumn{6}{c}{\text{Transitions}} & u_r\left(m_p\right/m_e) & u_r\left(m_d\right/m_e) & u_r\left(m_t\right/m_e) & u_r\left(R_{\infty }\right) & u\left(r_p\right) & u\left(r_d\right) & u\left(r_t\right) \\
 \text{} & \text{HD}^+ & \text{H}_2^+ & \text{D}_2^+ & \text{HT}^+ & \text{DT}^+ & \text{T}_2^+ & \left(10^{-12}\right) & \left(10^{-12}\right) & \left(10^{-12}\right) & \left(10^{-12}\right) & \text{(am)} & \text{(am)} & \text{(am)} \\
\hline
 5 & 4,{\bf 6} & \text{} & 13 & 18 & \text{} & 23 & 9.7 & 14. & 4.6 & 29. & 31. & 12. & 14. \\
 6 & 2,4,{\bf 6} & \text{} & 13 & 18 & \text{} & 23 & 0.49 & 1.2 & 0.72 & 14. & 15. & 5.9 & 7.0 \\
 7 & 2,4,{\bf 6} & \text{} & 13 & 18 & \text{} & \text{23,24} & 0.18 & 0.36 & 0.23 & 1.2 & 1.4 & 0.54 & 0.66 \\
 8 & 2,4,{\bf 6} & 9 & 13 & 18 & \text{} & \text{23,24} & 0.12 & 0.27 & 0.20 & 1.1 & 1.3 & 0.51 & 0.61 \\
 9 & 2,4,{\bf 6} & 9 & 13 & \text{18,19} & \text{} & \text{23,24} & 0.12 & 0.27 & 0.20 & 1.1 & 1.3 & 0.51 & 0.61 \\
\hline
 \multicolumn{7}{c}{\text{CODATA 2022}} & 17 & 17 & 38 & 1.1 & 0.64 & 0.27 & 86 \\
\hline
\end{array}
\]
     \caption{Similar to Tab.~\ref{tab:T no high-v}, but including one ``high-$v$'' transition, excluding pure rotational transitions as well as \DTplus.}
     \label{tab:T easy A}
\end{table*}
%\clearpage
\begin{table*}[h]
\footnotesize
     \[
\begin{array}{cccccccccccccc}
\hline
 n & \multicolumn{6}{c}{\text{Transitions}} & u_r\left(m_p\right/m_e) & u_r\left(m_d\right/m_e) & u_r\left(m_t\right/m_e) & u_r\left(R_{\infty }\right) & u\left(r_p\right) & u\left(r_d\right) & u\left(r_t\right) \\
 \text{} & \text{HD}^+ & \text{H}_2^+ & \text{D}_2^+ & \text{HT}^+ & \text{DT}^+ & \text{T}_2^+ & \left(10^{-12}\right) & \left(10^{-12}\right) & \left(10^{-12}\right) & \left(10^{-12}\right) & \text{(am)} & \text{(am)} & \text{(am)} \\
\hline
 5 & 3 & {\bf 12} & 13 & 19 & 22 & \text{} & 10. & 11. & 1.3 & 27. & 29. & 11. & 13. \\
 6 & \text{1,3} & {\bf 12} & 13 & 19 & 22 & \text{} & 5.3 & 4.4 & 1.2 & 2.1 & 2.3 & 0.91 & 1.4 \\
 7 & \text{1,3} & {\bf 12} & 13 & 19 & 22 & 23 & 0.29 & 0.65 & 0.52 & 1.2 & 1.4 & 0.56 & 0.67 \\
 8 & \text{1,3} & 9,{\bf 12} & 13 & 19 & 22 & 23 & 0.23 & 0.37 & 0.33 & 1.2 & 1.4 & 0.55 & 0.65 \\
 9 & \text{1,3} & 9,{\bf 12} & 13 & 19 & 22 & \text{23,24} & 0.23 & 0.33 & 0.33 & 0.94 & 1.1 & 0.45 & 0.54 \\
 10 & \text{1,2,3} & 9,{\bf 12} & 13 & 19 & 22 & \text{23,24} & 0.21 & 0.30 & 0.29 & 0.88 & 1.1 & 0.43 & 0.52 \\
 11 & \text{1,2,3,4} & 9,{\bf 12} & 13 & 19 & 22 & \text{23,24} & 0.21 & 0.30 & 0.28 & 0.87 & 1.1 & 0.43 & 0.51 \\
\hline
 \multicolumn{7}{c}{\text{CODATA 2022}} & 17 & 17 & 38 & 1.1 & 0.64 & 0.27 & 86 \\
\hline
\end{array}
\]
     \caption{Similar to Tab.~\ref{tab:T no high-v}, with a selection of experimentally convenient transitions.
     }
     \label{tab:T easy B}
\end{table*}

\begin{table*}[h]
\footnotesize
     \[
\begin{array}{cccccccccccccc}
\hline
 n & \multicolumn{6}{c}{\text{Transitions}} & u_r\left(m_p\right/m_e) & u_r\left(m_d\right/m_e) & u_r\left(m_t\right/m_e) & u_r\left(R_{\infty }\right) & u\left(r_p\right) & u\left(r_d\right) & u\left(r_t\right) \\
 \text{} & \text{HD}^+ & \text{H}_2^+ & \text{D}_2^+ & \text{HT}^+ & \text{DT}^+ & \text{T}_2^+ & \left(10^{-12}\right) & \left(10^{-12}\right) & \left(10^{-12}\right) & \left(10^{-12}\right) & \text{(am)} & \text{(am)} & \text{(am)} \\
\hline
 5 & 4,{\bf 6} & \text{} & 13 & 18 & \text{} & 23 & 9.7 & 14. & 4.6 & 29. & 31. & 12. & 14. \\
 6 & 1,4,{\bf 6} & \text{} & 13 & 18 & \text{} & 23 & 4.2 & 4.4 & 1.2 & 3.0 & 3.2 & 1.3 & 2.0 \\
 7 & 1,2,4,{\bf 6} & \text{} & 13 & 18 & \text{} & 23 & 0.19 & 0.35 & 0.22 & 1.2 & 1.5 & 0.57 & 0.69 \\
 8 & 1,2,4,{\bf 6} & 9 & 13 & 18 & \text{} & 23 & 0.14 & 0.24 & 0.17 & 1.2 & 1.4 & 0.56 & 0.67 \\
 9 & 1,2,3,4,{\bf 6} & 9 & 13 & 18 & \text{} & 23 & 0.14 & 0.23 & 0.17 & 1.2 & 1.4 & 0.55 & 0.65 \\
\hline
 \multicolumn{7}{c}{\text{CODATA 2022}} & 17 & 17 & 38 & 1.1 & 0.64 & 0.27 & 86 \\
\hline
\end{array}
\]
     \caption{Similar to Tab.~\ref{tab:T no high-v}, but including only one ``high-$v$'' transition and two transitions in tritium-containing isotopologues.}
     \label{tab:T 2 TMI 1 high-v}
\end{table*}

%%%%%%%%%%%%%%%%%%%%%%%%%%%%%%%%%%%%%%%%%%%%%%%%%
\vfill\clearpage

%%%%%%%%%%%%%%%%%%%%%%%%%%%%%%%%

\clearpage
\section{Example with larger values of $u_0$}
\label{Appendix: example with larger value of Bethe log uncertainty}

We study the impact of numerical uncertainties on the scenarios presented in Tab.~\ref{tab:no-T without high-v}-c (no tritium-containing MHI) and \ref{tab:T best B} (T-MHI). The values $u_0^{\text{proj}}=10^{-10}$, see Tab.~\ref{tab:no-T best large1}, \ref{tab:T best large1 B} and $u_0^{\text{today}}=10^{-9}$, see Tab.~\ref{tab:no-T best large2}, \ref{tab:T best large2 B}, are considered. The latter value corresponds to the current numerical precision of the Bethe logarithm.

\begin{table*}[h!]
\footnotesize
     \[
\begin{array}{ccccccccc}
\hline
 n & \multicolumn{3}{c}{\text{Transitions}} & u_r\left(\mu _{pd}\right/m_e) & u_r\left(m_p\right/m_e) & u_r\left(R_{\infty }\right) & u\left(r_p\right) & u\left(r_d\right) \\
 \text{} & \text{HD}^+ & \text{H}_2^+ & \text{D}_2^+ & \left(10^{-12}\right) & \left(10^{-12}\right) & \left(10^{-12}\right) & \text{(am)} & \text{(am)} \\
\hline
 3 & 1,{\bf 5} & 9 & \text{} & 11. & 2.9 & 16. & 17. & 6.9 \\
 4 & 1,2,{\bf 5} & 9 & \text{} & 3.0 & 2.5 & 1.8 & 2.0 & 0.80 \\
 5 & 1,2,3,{\bf 5} & 9 & \text{} & 0.34 & 0.46 & 1.5 & 1.7 & 0.66 \\
 6 & 1,2,3,{\bf 5} & 9 & 13 & 0.34 & 0.37 & 1.4 & 1.6 & 0.64 \\
 7 & 1,2,3,{\bf 5,8} & 9 & 13 & 0.34 & 0.35 & 1.4 & 1.6 & 0.62 \\
 8 & 1,2,3,{\bf 5,6,8} & 9 & 13 & 0.33 & 0.34 & 1.4 & 1.5 & 0.61 \\
 9 & 1,2,3,4,{\bf 5,6,8} & 9 & 13 & 0.31 & 0.33 & 1.3 & 1.5 & 0.61 \\
\hline
\multicolumn{4}{c}{\text{CODATA 2022}} & 17 & 17 & 1.1 & 0.64 & 0.27 \\
\hline
\end{array}
\]
     \caption{Similar to Tab.~\ref{tab:no-T without high-v}-c, but assuming $u_0^{\text{proj}} = 10^{-10}$.}
     \label{tab:no-T best large1}
\end{table*}

\begin{table*}[h!]
\footnotesize
     \[
\begin{array}{ccccccccc}
\hline
 n & \multicolumn{3}{c}{\text{Transitions}} & u_r\left(\mu _{pd}\right/m_e) & u_r\left(m_p\right/m_e) & u_r\left(R_{\infty }\right) & u\left(r_p\right) & u\left(r_d\right) \\
 \text{} & \text{HD}^+ & \text{H}_2^+ & \text{D}_2^+ & \left(10^{-12}\right) & \left(10^{-12}\right) & \left(10^{-12}\right) & \text{(am)} & \text{(am)} \\
\hline
 3 & 1,{\bf 5} & 9 & \text{} & 12. & 4.3 & 17. & 18. & 7.0 \\
 4 & 1,3,{\bf 5} & 9 & \text{} & 3.3 & 4.1 & 7.0 & 7.4 & 2.9 \\
 5 & 1,2,3,{\bf 5} & 9 & \text{} & 3.0 & 3.3 & 6.9 & 7.4 & 2.9 \\
 6 & 1,2,3,{\bf 5} & \text{9,11} & \text{} & 2.9 & 3.1 & 6.9 & 7.3 & 2.9 \\
 7 & 1,2,3,{\bf 5} & \text{9,11} & 13 & 2.9 & 3.1 & 6.8 & 7.2 & 2.9 \\
 8 & 1,2,3,4,{\bf 5} & \text{9,11} & 13 & 2.9 & 3.0 & 6.8 & 7.2 & 2.9 \\
 9 & 1,2,3,4,{\bf 5,6} & \text{9,11} & 13 & 2.7 & 2.9 & 6.7 & 7.2 & 2.8 \\
\hline
\multicolumn{4}{c}{\text{CODATA 2022}} & 17 & 17 & 1.1 & 0.64 & 0.27 \\
\hline
\end{array}
\]
     \caption{Similar to Tab.~\ref{tab:no-T without high-v}-c, but assuming $u_0^{\text{today}} = 10^{-9}$.}
     \label{tab:no-T best large2}
\end{table*}

\begin{table*}[h!]
\footnotesize
     \[
\begin{array}{cccccccccccccc}
\hline
 n & \multicolumn{6}{c}{\text{Transitions}} & u_r\left(m_p\right/m_e) & u_r\left(m_d\right/m_e) & u_r\left(m_t\right/m_e) & u_r\left(R_{\infty }\right) & u\left(r_p\right) & u\left(r_d\right) & u\left(r_t\right) \\
 \text{} & \text{HD}^+ & \text{H}_2^+ & \text{D}_2^+ & \text{HT}^+ & \text{DT}^+ & \text{T}_2^+ & \left(10^{-12}\right) & \left(10^{-12}\right) & \left(10^{-12}\right) & \left(10^{-12}\right) & \text{(am)} & \text{(am)} & \text{(am)} \\
\hline
 5 & 1,{\bf 5} & 9 & \text{} & 18 & \text{} & 24 & 2.9 & 36. & 1.1 & 16. & 17. & 6.9 & 8.7 \\
 6 & 1,{\bf 5} & 9 & 13 & 18 & \text{} & 24 & 2.6 & 1.2 & 0.93 & 2.8 & 3.1 & 1.2 & 1.4 \\
 7 & 1,{\bf 5} & 9 & 13 & 18 & 21 & 24 & 0.85 & 0.63 & 0.80 & 1.2 & 1.4 & 0.56 & 0.72 \\
 8 & 1,3,{\bf 5} & 9 & 13 & 18 & 21 & 24 & 0.49 & 0.61 & 0.56 & 1.1 & 1.3 & 0.52 & 0.63 \\
 9 & 1,3,{\bf 5} & 9 & 13 & \text{17,18} & 21 & 24 & 0.40 & 0.49 & 0.54 & 1.0 & 1.2 & 0.49 & 0.60 \\
 10 & 1,3,{\bf 5} & 9 & 13 & \text{17,18} & 21 & \text{23,24} & 0.37 & 0.47 & 0.49 & 0.97 & 1.2 & 0.46 & 0.57 \\
 11 & 1,3,4,{\bf 5} & 9 & 13 & \text{17,18} & 21 & \text{23,24} & 0.34 & 0.47 & 0.47 & 0.97 & 1.2 & 0.46 & 0.57 \\
 12 & 1,3,4,{\bf 5} & 9 & 13 & \text{17,18,19} & 21 & \text{23,24} & 0.34 & 0.46 & 0.45 & 0.97 & 1.2 & 0.46 & 0.56 \\
\hline
 \multicolumn{7}{c}{\text{CODATA 2022}} & 17 & 17 & 38 & 1.1 & 0.64 & 0.27 & 86 \\
\hline
\end{array}
\]
     \caption{Similar to Tab.~\ref{tab:T best B}, but assuming $u_0^{\text{proj}} = 10^{-10}$. 
     }
     \label{tab:T best large1 B}
\end{table*}

\begin{table*}[h!]
\footnotesize
     \[
\begin{array}{cccccccccccccc}
\hline
 n & \multicolumn{6}{c}{\text{Transitions}} & u_r\left(m_p\right/m_e) & u_r\left(m_d\right/m_e) & u_r\left(m_t\right/m_e) & u_r\left(R_{\infty }\right) & u\left(r_p\right) & u\left(r_d\right) & u\left(r_t\right) \\
 \text{} & \text{HD}^+ & \text{H}_2^+ & \text{D}_2^+ & \text{HT}^+ & \text{DT}^+ & \text{T}_2^+ & \left(10^{-12}\right) & \left(10^{-12}\right) & \left(10^{-12}\right) & \left(10^{-12}\right) & \text{(am)} & \text{(am)} & \text{(am)} \\
\hline
 5 & 1,{\bf 5} & 9 & \text{} & 18 & \text{} & 24 & 4.3 & 37. & 5.5 & 17. & 18. & 7.0 & 8.8 \\
 6 & 1,3,{\bf 5} & 9 & \text{} & 18 & \text{} & 24 & 4.1 & 3.6 & 5.5 & 7.0 & 7.4 & 2.9 & 3.7 \\
 7 & 1,3,{\bf 5} & 9 & \text{} & 18 & 21 & 24 & 4.0 & 3.6 & 5.5 & 5.7 & 6.1 & 2.4 & 3.1 \\
 8 & 1,3,{\bf 5} & 9 & \text{} & \text{17,18} & 21 & 24 & 3.3 & 3.6 & 5.2 & 5.0 & 5.4 & 2.1 & 2.7 \\
 9 & 1,3,{\bf 5} & 9 & \text{} & \text{17,18} & 21 & \text{23,24} & 3.3 & 3.6 & 4.8 & 4.9 & 5.3 & 2.1 & 2.6 \\
 10 & 1,3,{\bf 5} & 9 & 13 & \text{17,18} & 21 & \text{23,24} & 3.0 & 2.8 & 4.3 & 4.9 & 5.3 & 2.1 & 2.6 \\
 11 & 1,3,{\bf 5} & 9 & 13 & \text{17,18,19} & 21 & \text{23,24} & 3.0 & 2.8 & 4.1 & 4.9 & 5.3 & 2.1 & 2.6 \\
 11 & 1,3,{\bf 5} & 9 & 13 & \text{17,18,19,20} & 21 & \text{23,24} & 2.9 & 2.8 & 3.9 & 4.9 & 5.3 & 2.1 & 2.6 \\
\hline
 \multicolumn{7}{c}{\text{CODATA 2022}} & 17 & 17 & 38 & 1.1 & 0.64 & 0.27 & 86 \\
\hline
\end{array}
\]
     \caption{Similar to Tab.~\ref{tab:T best B}, but assuming $u_0^{\text{today}} = 10^{-9}$. 
     }
     \label{tab:T best large2 B}
\end{table*}

%\clearpage
\section{Examples with smaller value of the QED uncertainty}
\label{Appendix: examples with smaller value of the QED uncertainty}

We study the impact of reducing QED uncertainties on the scenarios presented in Tab.~\ref{tab:no-T without high-v}-c (no tritium-containing MHI), see Tab.~\ref{tab:no-T best red}, and \ref{tab:T best B} (T-MHI), see Tab.~\ref{tab:T best red B}.

\begin{table*}[h!]
\footnotesize
     \[
\begin{array}{ccccccccc}
\hline
 n & \multicolumn{3}{c}{\text{Transitions}} & u_r\left(\mu _{pd}\right/m_e) & u_r\left(m_p\right/m_e) & u_r\left(R_{\infty }\right) & u\left(r_p\right) & u\left(r_d\right) \\
 \text{} & \text{HD}^+ & \text{H}_2^+ & \text{D}_2^+ & \left(10^{-12}\right) & \left(10^{-12}\right) & \left(10^{-12}\right) & \text{(am)} & \text{(am)} \\
\hline
 3 & 1,{\bf 5} & 9 & \text{} & 1.8 & 0.59 & 2.6 & 2.5 & 0.98 \\
 4 & 1,4,{\bf 5} & 9 & \text{} & 0.20 & 0.51 & 1.1 & 1.2 & 0.46 \\
 5 & 1,4,{\bf 5} & 9 & 13 & 0.15 & 0.11 & 1.1 & 1.2 & 0.46 \\
 6 & 1,4,{\bf 5},{\bf 6} & 9 & 13 & 0.15 & 0.073 & 0.88 & 0.90 & 0.38 \\
 7 & 1,2,4,{\bf 5},{\bf 6} & 9 & 13 & 0.10 & 0.072 & 0.70 & 0.78 & 0.32 \\
 8 & 1,2,4,{\bf 5},{\bf 6} & 9 & 13,{\bf 14} & 0.095 & 0.071 & 0.63 & 0.74 & 0.30 \\
 9 & 1,2,4,{\bf 5},{\bf 6},{\bf 8} & 9 & 13,{\bf 14} & 0.094 & 0.071 & 0.59 & 0.71 & 0.29 \\
\hline
\multicolumn{4}{c}{\text{CODATA 2022}} & 17 & 17 & 1.1 & 0.64 & 0.27 \\
\hline
\end{array}
\]
     \caption{Similar to Tab.~\ref{tab:no-T without high-v}c, but assuming reduction of QED uncertainties by a factor of 8. 
     }
     \label{tab:no-T best red}
\end{table*}

\begin{table*}[h!]
\footnotesize
     \[
\begin{array}{cccccccccccccc}
\hline
 n & \multicolumn{6}{c}{\text{Transitions}} & u_r\left(m_p\right/m_e) & u_r\left(m_d\right/m_e) & u_r\left(m_t\right/m_e) & u_r\left(R_{\infty }\right) & u\left(r_p\right) & u\left(r_d\right) & u\left(r_t\right) \\
 \text{} & \text{HD}^+ & \text{H}_2^+ & \text{D}_2^+ & \text{HT}^+ & \text{DT}^+ & \text{T}_2^+ & \left(10^{-12}\right) & \left(10^{-12}\right) & \left(10^{-12}\right) & \left(10^{-12}\right) & \text{(am)} & \text{(am)} & \text{(am)} \\
\hline
 5 & 4,{\bf 8} & \text{} & 13 & 18 & \text{} & 23 & 2.2 & 2.8 & 1.6 & 5.1 & 5.2 & 2.0 & 2.3 \\
 6 & 2,4,{\bf 8} & \text{} & 13 & 18 & \text{} & 23 & 0.43 & 0.34 & 0.33 & 1.7 & 1.6 & 0.64 & 0.76 \\
 7 & 2,4,{\bf 5},{\bf 8} & \text{} & 13 & 18 & \text{} & 23 & 0.13 & 0.30 & 0.21 & 0.84 & 0.95 & 0.37 & 0.44 \\
 8 & 1,2,4,{\bf 5},{\bf 8} & \text{} & 13 & 18 & \text{} & 23 & 0.12 & 0.28 & 0.20 & 0.69 & 0.83 & 0.32 & 0.38 \\
 9 & 1,2,4,{\bf 5},{\bf 8} & 9 & 13 & 18 & \text{} & 23 & 0.075 & 0.23 & 0.17 & 0.67 & 0.76 & 0.31 & 0.37 \\
 10 & 1,2,4,{\bf 5},{\bf 8} & 9 & 13,{\bf 14} & 18 & \text{} & 23 & 0.074 & 0.20 & 0.16 & 0.61 & 0.72 & 0.29 & 0.35 \\
 11 & 1,2,4,{\bf 5},{\bf 8} & 9 & 13,{\bf 14} & 18 & 21 & 23 & 0.072 & 0.19 & 0.15 & 0.57 & 0.69 & 0.28 & 0.34 \\
 12 & 1,2,4,{\bf 5},{\bf 8} & 9 & 13,{\bf 14} & \text{17,18} & 21 & 23 & 0.071 & 0.19 & 0.15 & 0.54 & 0.67 & 0.27 & 0.33 \\
\hline
 \multicolumn{7}{c}{\text{CODATA 2022}} & 17 & 17 & 38 & 1.1 & 0.64 & 0.27 & 86 \\
\hline
\end{array}
\]
     \caption{Similar to Tab.~\ref{tab:T best B}, but assuming reduction of QED uncertainties by a factor of 8. 
     }
     \label{tab:T best red B}
\end{table*}

%%%%%%%%%%%%%%%%%%%%%%%%%%%%%%%%%%%%%%%%%%%%%%%%%

\section{Effect of imperfect correlations between theoretical uncertainties}
\label{Appendix - Imperfect correlations}

The strong correlations between theoretical uncertainties are one of the key ingredient allowing for precise determination of FCs from a set of measurements in MHIs. In our analysis, we assumed that the uncertainties of two different transition frequencies $f_i$, $f_j$ stemming from a given QED term, $u_{\text{QED},k}(\delta f_i^{\rm theor})$ and $u_{\text{QED},k}(\delta f_j^{\rm theor})$ are perfectly correlated to each other (see End Matter, Eq.~\ref{eq:cov matrix}). Here, we briefly discuss this hypothesis and explore the effect of possible deviations from it.

We recall that for the evaluation of uncertainties and correlations, we follow Ref.~\cite{Korobov2021}. In that work, uncalculated QED terms are estimated by approximating them by a delta-function effective potential, the prefactor of which is equal to that for the ground state of a hydrogen atom. The associated uncertainty taken as equal to 100\% of the contribution from the estimated term to the transition frequencies. This implies the following: the closer the actual potential describing these contributions is to a delta-function term, the stronger the correlations between the uncertainties will be.

The two largest sources of uncertainty are the contributions of order $m_e \alpha^8$ and higher to the one-loop self-energy and two-loop corrections. Being corrections to the bound electron, these two terms are expected to behave similarly to the atomic case, i.e. to be only weakly state-dependent and well approximated by a delta-function effective potential. The assumption of perfect (or very close to perfect) correlation is thus well motivated for these terms.

However, for two other (much smaller) sources of uncertainty, substantial deviations from the case of perfect correlation could exist. One of them is the uncertainty associated with the use of the adiabatic approximation, in particular for calculation of the $m_e \alpha^6$-order relativistic correction. The other one is the uncertainty stemming from the relativistic-recoil correction of order $m_e \alpha^6$. Being related to nuclear motion, both types of corrections may have a different dependence on ro-vibrational levels with respect to their estimated form, which would imply a reduction of the correlation. It is thus important to explore the consequences of imperfect correlations for these two contributions.

We performed additional simulations in the scenario described in the last line of Tab.~\ref{tab:T best B}, where we assumed a correlation coefficient $r_{k,ij}=0.99$ and $r_{k,ij}=0.95$ (see End Matter, Eq.~\ref{eq:cov matrix}) for the above-mentioned contributions to the theoretical uncertainty, see Tab.~\ref{tab:T best imperfect correlation}. 

A significant degradation of the results is observed, especially in the mass ratios.
In conclusion, further theoretical work is required to refine the estimate of correlations. Most desirable would be to calculate (i) the $m_e \alpha^6$-order relativistic correction in a full three-body approach, and (ii) the $m_e \alpha^6$-order relativistic-recoil correction, which would eliminate these two potentially problematic sources of uncertainty. Fortunately, both calculations appear less challenging than that of the $m_e \alpha^8$-order terms that limit the overall uncertainty.

\begin{table*}[!]
\footnotesize
$\begin{array}{cccccccccccccc}
\hline
 r & \multicolumn{6}{c}{\text{Transitions}} & u_r\left(m_p\right/m_e) & u_r\left(m_d\right/m_e) & u_r\left(m_t\right/m_e) & u_r\left(R_{\infty }\right) & u\left(r_p\right) & u\left(r_d\right) & u\left(r_t\right) \\
 \text{} & \text{HD}^+ & \text{H}_2^+ & \text{D}_2^+ & \text{HT}^+ & \text{DT}^+ & \text{T}_2^+ & \left(10^{-12}\right) & \left(10^{-12}\right) & \left(10^{-12}\right) & \left(10^{-12}\right) & \text{(am)} & \text{(am)} & \text{(am)} \\
\hline
 1.00 & 1,2,4,{\bf 5},{\bf 8} & 9 & 13,{\bf 14} & \text{17,18} & 21 & 23 & 0.071 & 0.19 & 0.15 & 0.55 & 0.82 & 0.32 & 0.39 \\
 0.99 & 1,2,4,{\bf 5},{\bf 8} & 9 & 13,{\bf 14} & \text{17,18} & 21 & 23 & 0.48 & 0.66 & 0.57 & 0.92 & 1.1 & 0.45 & 0.55 \\
 0.95 & 1,2,4,{\bf 5},{\bf 8} & 9 & 13,{\bf 14} & \text{17,18} & 21 & 23 & 0.98 & 1.1 & 1.1 & 1.4 & 1.6 & 0.64 & 0.80 \\
\hline
 \multicolumn{7}{c}{\text{CODATA 2022}} & 17 & 17 & 38 & 1.1 & 0.64 & 0.27 & 86 \\
\hline
\end{array}
$
\caption{Similar to the last line of Tab.~\ref{tab:T best B} (with $n=12$ transitions), but with different values of the correlation coefficient $r_{k,ij}$ attributed to uncertainties associated with the adiabatic approximation and with the $m_e \alpha^6$-order relativistic-recoil correction.} \label{tab:T best imperfect correlation}
\end{table*}

\clearpage
\newpage

\section{Spectroscopy for the determination of $r_t$ and $m_t$}

\label{Appendix - Spectroscopy for rt}

\subsection{T-MHI spectroscopy}
\label{Appendix - T-MHI spectroscopy}
The masses of T-MHI are similar to that of \HDplus, therefore sympathetic cooling by $^9$Be$^+$ will be possible. 
The nuclear spin of the triton is the same as that of the proton, $I=1/2$, and the magnetic moments are similar. Therefore, the spin structure of the levels and the transition dipole and transition quadrupole moments of \T2plus and \DTplus\ are similar to those for \H2plus and \HDplus, respectively - for both of which spectroscopy has already been demonstrated. For \HTplus\ the spectroscopy may turn out to be simpler (more efficient) than that of \HDplus and \DTplus, due to the smaller number of spin states in the former, a fact that simplifies state preparation. The hyperfine structure theory developed and tested on the non-radioactive MHI has already been extended to \HTplus \cite{Bekbaev2013}.

\subsection{Atomic tritium spectroscopy}
\label{Appendix - Atomic T spectroscopy}
The T-REX project \cite{Schmidt2018} aims at developing two-photon 1s-2s spectroscopy in T and to extract $r_t$ in a similar manner as done for H and D. A first-generation experiment shall be performed using a discharge in a cell containing  warm T$_2$ gas. Later, with sympathetically cooled atomic tritium gas, high accuracy is targeted, a 300-fold reduction of $u(r_t)$. The proposal suggests creating and trapping a cold cloud of tritium atoms as follows. A cryogenic source produces a cold atomic tritium beam. A magnetic multipole guide then selects the low-velocity portion of the beam and directs these atoms into a magnetic trap with magnetic minimum of 0.4\,Tesla. There, the trapped T atoms will interact with a dense ensemble of laser-cooled lithium atoms, that acts as a cold target to cool the tritium atoms. Once the T atoms are sufficiently cooled, the lithium atoms are removed from the trap, leaving a cold T sample for the precision measurement. 

\clearpage

\end{document}